\pdfoutput=1 
\documentclass[aps,prd,nofootinbib,floatfix,superscriptaddress,twocolumn]{revtex4}

\usepackage{graphicx}
\usepackage{bm}
\usepackage{amsmath}
\usepackage{amssymb}

\usepackage[colorlinks,pdfstartview=FitH]{hyperref}
\hypersetup{linkcolor=blue,citecolor=blue,filecolor=black,urlcolor=blue}

\graphicspath{{Figures/}}

\newcommand\Th{T_{\rm hf}}
\newcommand\chih{\chi_{\rm hf}}
\newcommand\Tkf{T_{\rm kf}}
\newcommand\Nnet{N_{\rm net}}
\newcommand\tauf{{\tau_{\rm f}}}
\newcommand\f{{\rm f}}

\begin{document}


\title[]{Hydrodynamics of charge fluctuations and balance functions}
\author{Bo Ling}
\affiliation{
Physics Department,  University of Illinois at Chicago,
Chicago, IL, 60607%
}
\author{Todd Springer}
\affiliation{
Physics Department,  University of Illinois at Chicago,
Chicago, IL, 60607%
}
\affiliation{
Department of Physics and Astronomy,
Colgate University, Hamilton, NY 13346
}
\author{Mikhail Stephanov}
\affiliation{
Physics Department,  University of Illinois at Chicago,
Chicago, IL, 60607%
}%
\affiliation{
 Enrico Fermi Institute, University of Chicago,
Chicago, Illinois 60637}


\date{June 3, 2014}

\begin{abstract}
  We apply stochastic hydrodynamics to the study of charge density
  fluctuations in QCD matter undergoing Bjorken expansion. We find
  that the charge density correlations are given by a time integral
  over the history of the system, with the dominant contribution
  coming from the QCD crossover region where the change of
  susceptibility per entropy, $\chi T/s$, is most significant.  We
  study the rapidity and azimuthal angle dependence of the resulting
  charge balance function using a simple analytic model of heavy-ion
  collision evolution.  Our results are in agreement with experimental
  measurements, indicating that hydrodynamic fluctuations contribute
  significantly to the measured charge correlations in high energy
  heavy-ion collisions. The sensitivity of the balance function to the
  value of the charge diffusion coefficient $D$ allows us to estimate
  the typical value of this coefficient in the crossover region to be
  rather small, of the order of $(2 \pi T)^{-1}$, characteristic of a
  strongly coupled plasma.
\end{abstract}

\maketitle

\section{Introduction}
Event-by-event fluctuations and two particle
correlations~\cite{Jeon:2003gk} in high-energy heavy-ion collision
experiments provide valuable information about the collective
dynamics: thermal and transport properties of the hot and dense QCD
matter.  Much recent effort has been devoted to the measurement and
understanding of the correlations observed at RHIC and LHC. In this paper
we focus on charge-dependent correlations.  The suppression of the
event-by-event fluctuations of the net (electric) charge has been
proposed as a signature of the QGP formation
\cite{Jeon:2000wg,Asakawa:2000wh} and studied
experimentally~\cite{Abelev:2008jg, Abelev:2012pv}.  More differential
measures of charge fluctuations, such as the azimuthal and rapidity
dependence of charge-dependent correlations have also attracted
interest. The difference between like-sign and unlike-sign
correlations, often expressed as balance functions
\cite{Bass:2000az,Jeon:2001ue,Bozek:2012en,Pratt:2012dz}, have also been
studied experimentally
\cite{Abelev:2009jv,Timmins:2011um,Aggarwal:2010ya,Abelev:2013csa}.

In this paper, we apply relativistic stochastic hydrodynamics
\cite{Kapusta:2011gt} to study the balance functions in heavy-ion
collisions.  Hydrodynamic equations describe the evolution of
conserved quantities such as energy, momentum, and charge averaged
over a statistical thermodynamic ensemble. Fluctuations around 
  static equilibrium can be described using the fluctuation-dissipation
theorem. In order to describe fluctuations around a {\em non-static} solution
of the hydrodynamic equations (such as, e.g., Bjorken expansion) one can
introduce local noise into the hydrodynamic equations, as has been
proposed by Landau and Lifshitz \cite{LandauStatV9}, but has only recently been
applied in the context of relativistic heavy-ion collisions~\cite{Kapusta:2011gt}.
The hydrodynamic evolution of such local noise naturally leads to
observable particle correlations.  

As emphasized in Ref.~\cite{Kapusta:2011gt} (see also references therein),
the hydrodynamic fluctuations are not the only source of observed
correlations. Other sources include initial state fluctuations,
fluctuations induced by rare hard processes (jets) and final state
(freezeout) fluctuations. These contributions remain important in the case
of charge correlations. However, the fact that the inital state in the
ultra-relativistic heavy-ion collisions is dominated by saturated glue
carrying no electric charge might suppress the contribution of initial
state fluctuations to charge correlations we discuss here, compared to
entropy fluctuations discussed in Ref.~\cite{Kapusta:2011gt}.

Under the conditions of the boost-invariant $1+1$ dimensional Bjorken
expansion \cite{Bjorken:1982qr}, the effect of stochastic baryon
number current at nonzero mean baryon density was studied in
Ref.~\cite{Kapusta:2011gt} without considering diffusion. Here we consider
the effect of diffusion and a stochastic charge current at zero mean
charge density. The analytical simplicity of the Bjorken solution
allows us to understand in detail the mechanisms at work while using a
phenomenologically reasonable description of a heavy-ion collision.
As a step towards adequately addressing azimuthal dependence of
correlations we introduce transverse expansion on top of the Bjorken
solution using the standard ``blast wave'' model.

The paper is organized as follows: In Sec. \ref{FluctuationSection},
we briefly review hydrodynamics with noise.  We linearize the
stochastic hydrodynamic equations (around the Bjorken solution at zero
charge density) and analytically solve them.  We find the charge
correlations emerging as a superposition of contributions of past local
noise sources which have diffused over the time separating the source and the
observation. Successive contributions cancel each other unless the
quantity $\chi T/s$ (more precisely $\chi T\tau$) is changing with
time. Thus we find the dominant contribution coming from the crossover
region of the QCD phase diagram where the effective degrees of freedom
change from those of the quark-gluon plasma to those of the hadron gas.
In Sec. \ref{LatticeEOSSec}, we use the lattice QCD data
\cite{Borsanyi:2010cj, Borsanyi:2011sw}
to obtain the dependence of susceptibility per entropy $\chi T/s$ on
temperature which determines the magnitude of the charge correlations.
We apply these results to a simple semianalytical model of expansion
with Cooper-Frye freezeout and make an example comparison with
experimental data from RHIC in Section~\ref{FreezeOutSec}. We conclude
with a discussion in Sec. \ref{ConclusionSec}.

\section{Hydrodynamic Fluctuations}
\label{FluctuationSection}
\subsection{Hydrodynamics and noise}
Hydrodynamics describes the slow evolution of conserved
quantities such as energy, momentum, and conserved charges.  In
the case of QCD the charge could be the baryon number, electric
charge, or strangeness. At top energies at RHIC and at LHC most
particles in the final state are pions, which carry only electric
charge. Therefore in this work we shall focus on electric charge
fluctuations. The extension to other conserved charges
such as baryon number or strangeness should be straightforward.  The five
hydrodynamic equations of motion are the conservation equations for
energy-momentum and charge
\begin{eqnarray}
&&\nabla_\mu (T^{\mu\nu})=0,
\notag
\\
&&
\partial_\mu (\sqrt{-g}J^\mu)=0.
\label{ConservationEqns}
\end{eqnarray}
Here $\nabla_\mu$ denotes the covariant derivative with respect to the
background metric $g_{\mu \nu}$ and $g\equiv\det[g_{\mu\nu}]$  -- we shall
only consider flat space-time, but use curvilinear (Bjorken)
coordinates.  Fluctuations are described by adding stochastic noise
terms $S^{\mu\nu}$ and $I^{\mu}$, as explained in
\cite{LandauStatV5} or, in relativistic context, in
\cite{Kapusta:2011gt}:
\begin{eqnarray}
T^{\mu\nu} &=& T^{\mu\nu}_{\rm ideal} + \Delta T^{\mu\nu} +
S^{\mu\nu},
\notag
   \\
J^{\mu} &=& n u^{\mu} + \Delta J^{\mu} +
I^{\mu}.
\label{EnergyMomentumTensor}
\end{eqnarray}
Here, $n$ and $u^\mu$ are the charge density and fluid velocity, $T^{\mu \nu}_{\rm ideal}$ is the 
stress-energy tensor for an ideal fluid, and $\Delta T^{\mu\nu}$ and $\Delta J^\mu$ are dissipative (gradient)
corrections to stress and current.  The
dissipative correction to the current to the first order in gradients is given by
\begin{eqnarray}
\Delta J^{\mu}=\sigma T\Delta^\mu \left(\frac{\mu}{T}\right)\,,
\label{Jviscous}
\end{eqnarray}
where $\sigma$ is the charge conductivity, $\mu$ is the chemical
potential, and $\Delta^\mu \equiv \Delta^{\mu\nu}\partial_\nu$ ($\Delta^{\mu\nu} \equiv u^\mu u^\nu-g^{\mu\nu}$) is the spatial
derivative in the local rest frame of the fluid (whose 4-velocity is
$u^\mu$).
The diffusion coefficient $D$ is related to the conductivity by the
Einstein relation
\begin{eqnarray}
  D=\frac{\sigma}{\chi},
  \label{EinsteinRelation}
\end{eqnarray}
where $\chi$ is the electric charge susceptibility
\begin{eqnarray}
  \chi \equiv \left(\frac{\partial n}{ \partial \mu} \right)_T.
\end{eqnarray}

The hydrodynamic equations (\ref{ConservationEqns}) are
non-linear. However, in the domain of applicability of hydrodynamics
these equations can be linearized \cite{LandauStatV9} in the
perturbations around a given solution of the (non-linear)
deterministic equations of motion, i.e., Eqs.~(\ref{ConservationEqns})
without noise.  Such a linearized approach is sufficient to study
two-point correlations which are the subject of this paper.

To find two-point correlation functions of hydrodynamic variables we
need to know the two-point correlation functions of the noise.
One-point functions vanish by definition of the noise.  The
fluctuation-dissipation theorem determines the magnitude of the
two-point correlator:
\begin{eqnarray}\label{eq:II}
\langle I^\mu(x) \rangle&=&0, \\\label{eq:II-2}
\langle I^\mu(x)I^\nu(y)\rangle&=&2\sigma T\Delta^{\mu\nu}\delta(x-y),
\end{eqnarray}
where $\sigma$ and $T$ are functions of 
space and time given by the solution of the deterministic
(without noise) hydrodynamic Eqs.~(\ref{ConservationEqns}). 
The correlators of $S^{\mu \nu}$  are written down in
\cite{Kapusta:2011gt}, but we will not need them in this work.

Generalization to non-linear treatment of fluctuations is an
interesting problem, potentially relevant for the study of
higher-point correlations or fluctuations near a critical
point. Although linearized treatment is sufficient for the purposes of
this paper, it is worth keeping in mind the issues involved in the
non-linear generalization. The most obvious issue is that the noise
would become multiplicative since the magnitude of its correlator in
Eq.~(\ref{eq:II-2}) would be a function of the fluctuating
hydrodynamic variables. The formal definition in Eq.~(\ref{eq:II-2})
would have to be supplemented by a prescription (e.g., Ito or
Stratonovich) to resolve the well-known equal-time product ambiguity
(see, e.g., Ref.~\cite{GardinerStochastic}). The non-linearities also
give rise to short-distance singularities~\cite{Fox:1978} reminiscent
of the ultraviolet divergences in quantum field theories. Such issues
do not arise in the linearized treatment and we leave them outside of
the scope of this paper.

\subsection{Bjorken expansion and linear perturbations}
\label{BjorkenStochasticSec}

We shall use the well-known boost-invariant Bjorken solution
\cite{Bjorken:1982qr} of the
deterministic hydrodynamics Eqs.~(\ref{ConservationEqns}) as the
background for the linearized fluctuation analysis.
It is most convenient to describe the Bjorken
flow in the coordinates $(\tau,
\vec{x}_\perp, \eta)$ defined as
\begin{eqnarray}
  \tau &\equiv& \sqrt{t^2 - z^2}, \\
  \eta &\equiv& \tanh^{-1}\left(\frac{z}{t}\right).
\end{eqnarray}
The Bjorken time $\tau$ is invariant under boosts along the $z$ axis while
Bjorken rapidity $\eta$
shifts by a constant (the boost rapidity). The liquid undergoing
boost-invariant expansion is locally at rest
in these coordinates
\begin{equation}
\bar u^\mu(x) = \{1,\vec 0,0\}.
   \label{u1000}
 \end{equation}
while the energy (or entropy) density is a function of $\tau$, which can be found by solving
an ordinary differential equation.

We denote the background quantities with an overbar, and consider small perturbations to entropy density, flow velocity and
charge density expressed as $\rho \equiv \delta s / \bar{s}$, $\delta
u^x$, $\delta u^y$, $\delta u^\eta$, and $\delta n$:
\begin{eqnarray}
  \varepsilon(\tau,\vec{x}_\perp,\eta) &=& \bar{\varepsilon}(\tau) +
  \bar T(\tau) \bar{s}(\tau)\, \rho(\tau,\vec{x_\perp},\eta) 
\nonumber\\
&&+ \bar\mu(\tau)\, \delta n(\tau,\vec{x_\perp},\eta); \\
  u^\mu(\tau,\vec{x}_\perp, \eta) &=& \left\{1,\delta \vec{u}_\perp(\tau,\vec{x}_\perp,\eta), \delta u^\eta(\tau,\vec{x}_\perp,\eta)\right\};\\
  n(\tau, \vec{x}_\perp, \eta) &=& \bar{n}(\tau) + \delta n(\tau, \vec{x}_\perp, \eta);
\end{eqnarray}
where, as in Eq.~(\ref{u1000}), we are working in the Bjorken coordinates.  The quantity $\delta u^\tau$ vanishes at linear
order due to the unit norm constraint $u_\mu u^\mu = 1$.

In general, the fluctuations of the charge and the energy density mix
in Eq.(\ref{ConservationEqns}). However, in the special case of zero
background net charge density ($\bar{n} =0$) or, equivalently, zero
chemical potential ($\bar\mu=0$) the fluctuations of the charge
density $\delta n$ separate, at linear order considered here,
from the fluctuations of entropy density and flow velocity. Since we
are going to study only fluctuations of charge density, this
simplifies our task considerably. For top-energy RHIC collisions and
at LHC the chemical potential is very small compared to relevant
microscopic (QCD) scale and the approximation $\bar{\mu}=0$ is
useful. Since, as far as charge correlations are concerned, we can
ignore entropy and flow velocity fluctuations, we shall no longer
distinguish between quantities such as $\bar{s}$ and $s$, or $\bar T$
and $T$.

The stochastic charge diffusion equation in
Eq.(\ref{ConservationEqns}) becomes
\begin{eqnarray}
\partial_\tau J^\tau +\frac{J^\tau}{\tau}+\partial_\eta J^\eta+\vec{\nabla}_\perp \cdot\vec{J}_\perp=0
\label{BjorkenCurrentEqn}
\end{eqnarray}
Since for the fluid locally
at rest (\ref{u1000}) the only derivatives in
$\Delta^{\mu\nu}\nabla_\nu (\mu/T)$ are spatial and since $T$ depends on $\tau$ only, we can simplify
Eq. (\ref{Jviscous}) for $\Delta J^\mu$:  
\begin{eqnarray}
\Delta J^{\mu}= \sigma \Delta^\mu \mu = D \Delta^\mu n,
\end{eqnarray}
which is Fick's law of diffusion.  Substituting this into Eq. (\ref{BjorkenCurrentEqn}), we find
\begin{eqnarray}
\frac{1}{\tau} \partial_\tau \left(\tau \delta n \right) - D \left[\nabla_\perp^2 + \frac{1}{\tau^2}\partial_\eta^2 \right] \delta n  = -\nabla_i I^i - \nabla_\eta I^\eta.
\label{BjorkenDiffusion}
\end{eqnarray}
We use Latin indices $i,j,...$ to denote the two transverse
directions.

To facilitate the analysis of azimuthal correlations it is useful to
decompose the  noise current in the transverse plane
as \footnote{We are splitting a two-component vector into the gradient
  of a scalar, $I_S$, and a divergenceless two-vector, which, in two
  dimensions can also be written in terms of a scalar $I_V$.}
\begin{eqnarray}
  I^i = \tau \nabla_j \left[g^{ij}I_S - \epsilon^{ij}I_V \right].
\end{eqnarray}
Only $I_S$ will contribute to Eq.~(\ref{BjorkenDiffusion}). In order to solve
Eq.~(\ref{BjorkenDiffusion}), we express $\vec{x}_\perp$ in polar
coordinates $r$ and $\phi$, and use a Fourier-Bessel transformation for $\delta n$,
$I^\eta$ and $I_S$, which we define, for any function $f$, as
\begin{eqnarray}
f(\tau, r, \phi,\eta)&=& \sum_n \frac{e^{in\phi}}{2\pi}
\int_{-\infty}^\infty \frac{dk_\eta}{2\pi}e^{ik_\eta\cdot \eta}
\int_0^\infty dk_\perp k_\perp 
\nonumber\\&\times&
J_n(k_\perp r)\tilde{f}_n(\tau,k_\perp,k_\eta), \label{BesselTransform-f}\\
\tilde{f}_n(\tau,q_\perp,q_\eta) &=& \int_0^{2\pi} e^{-i n \phi} d\phi
\int_{-\infty}^{\infty} e^{-i q_\eta \eta} d \eta \int_0^{\infty} r dr
\nonumber\\&\times&
J_n(q_\perp r) f(\tau,r,\phi,\eta).
\label{BesselTransform}
\end{eqnarray}
Inverting the transformation requires the closure relation,
\begin{eqnarray}
  \int_0^{\infty} r dr J_n(k_\perp r) J_n(q_\perp r) = \frac{\delta(k_\perp - q_\perp)}{k_\perp}.
  \label{BesselClosure}
\end{eqnarray}

\subsection{Solution and correlations}
\label{sec:solut-corr}

The charge density fluctuation at a time $\tauf$ sourced by the
hydrodynamic noise $I$ is given by (upon Fourier-Bessel transform)
\begin{multline}\label{eq:delta-n}
\delta \tilde{n}_n (\tauf,k_\perp,k_\eta)
\\ \qquad
 = \frac{1}{\tauf}\int_{\tau_0}^{\tauf} d\tau \left[(\tau k_\perp)^2\tilde{I}_{S,n} - i \tau k_\eta \tilde{I}_n^\eta \right] e^{-H(\tauf,\tau,k_\perp, k_\eta)}
\end{multline}
where we defined
\begin{eqnarray}
H(\tauf,\tau, k_\perp,k_\eta) \equiv \int_{\tau}^\tauf d\tau^{\prime} D\left(\tau^{\prime}\right) \left(\frac{k_\eta^2}{{\tau^{\prime}}^2 }+ k_\perp^2\right).
\label{Hdef}
\end{eqnarray}
The coefficients of $k_\perp^2$ and $k_\eta^2$ in Eq.~(\ref{Hdef}) are the squared 
diffusion distances in the $x_\perp$ and $\eta$ directions. It is easy
to see that by considering the equation $dl^2=Ddt$ for the diffusion
(random walk) distance $dl$ in time $dt$ in a locally comoving frame.
The length element in Bjorken coordinates is
$dl^2=\tau^2d\eta^2+dx_\perp^2$ and the time element is
$dt=d\tau$. Thus the diffusion distance squared in the rapidity direction is
given by $(\Delta\eta)^2 = \int d\tau D/\tau^2$ and in the transverse
direction by $(\Delta x_\perp)^2= \int d\tau D$.

To determine the charge density correlations, one needs 
Fourier-Bessel transform of the noise correlators in Eq.~(\ref{eq:II}):
\begin{multline}
  \left<\tilde{I}_n^\eta(\tau, k_\perp, k_\eta)
    \tilde{I}_m^{*\eta}(\tau',q_\perp, q_\eta) \right> \\= \frac{2
    \sigma T}{\tau^3}  \delta(\tau-\tau')
  \hat{\delta}_{nm}(k,q), \label{eq:IIn}
\end{multline}
\begin{multline}
  \left<\tilde{I}_{S,n}(\tau, k_\perp, k_\eta)
    \tilde{I}_{S,m}^{*}(\tau',q_\perp, q_\eta) \right> \\
= \frac{2 \sigma T}{\tau^3 k_\perp^2}  \delta(\tau-\tau' ) \hat{\delta}_{nm}(k,q).\label{eq:IIS}
\end{multline}
Here we introduced a shorthand
\begin{eqnarray}
\hat{\delta}_{nm}(k,q) \equiv (2\pi)^2 \delta_{n,m}\delta(k_\eta-q_\eta)\frac{\delta(k_\perp -q_\perp)}{k_\perp}.
\label{deltahatdef}
\end{eqnarray}
Using Eqs.~(\ref{eq:delta-n}),~(\ref{eq:IIn}) and~(\ref{eq:IIS})  we find for the charge density correlations (at equal time $\tauf $)
\begin{multline}
  \left<\delta \tilde{n}_n(\tauf ,k_\perp,k_\eta) \delta  \tilde{n}^*_m(\tauf ,q_\perp, q_\eta) \right> = \hat{\delta}_{nm}(k,q)\\\times \frac{1}{\tauf ^2} \int_{\tau_0}^{\tauf } \frac{2 \sigma(\tau)T(\tau)}{\tau}
  \left[\tau^2 k_\perp^2 + k_\eta^2 \right]  e^{-2 H(\tauf ,\tau,k_\perp, k_\eta)} d\tau
  \label{nnsolution}
\end{multline}
With the aid of Eq. (\ref{EinsteinRelation}),
Eq. (\ref{nnsolution}) can be written as
\begin{multline}
    \left<\delta \tilde{n}_n(\tauf ,k_\perp,k_\eta) \delta
      \tilde{n}^*_m(\tauf ,q_\perp, q_\eta) \right> =
    \hat{\delta}_{nm}(k,q) \\\times \frac{1}{\tauf ^2} \int_{\tau_0}^{\tauf} \chi(\tau)T(\tau) \tau
  \frac{d}{d\tau} e^{-2 H(\tauf ,\tau,k_\perp, k_\eta)} d\tau.
\end{multline}
Finally, performing an integration by parts, we find
\begin{multline}
\Big\langle\delta\tilde{ n}_n(\tauf ,k_\perp,k_\eta)\delta
\tilde{n}^*_m(\tauf ,q_\perp,q_\eta)\Big\rangle =
{\hat{\delta}_{mn}(k,q)} \\\times \frac1{\tauf }\left[{\chi_\f  T_\f }- {s_\f }\tilde{A}(\tauf , k_\perp,k_\eta)\right]
\label{nnfinal}
\end{multline}
where we defined a dimensionless function in Fourier-Bessel space as
\begin{multline}
  \tilde{A}(\tauf , k_\perp, k_\eta)\equiv\frac{1}{s_\f \tauf
  }\Bigg(\chi_0 T_0\tau_0e^{-2H(\tauf ,\tau_0, k_\perp,k_\eta)}
\\+\int_{\tau_0}^{\tauf }d\tau e^{-2H(\tauf,\tau, k_\perp,k_\eta)}\frac{d(\chi T \tau)}{d\tau}\Bigg).
  \label{Adefinition}
\end{multline}

Note that
\begin{eqnarray}
  \tilde{A}(\tauf ,k_\perp=0,k_\eta=0)=\frac{\chi_\f T_\f }{s_\f }\,.
\end{eqnarray}
This, according to Eq.~(\ref{nnfinal}), implies the vanishing of
$\left<\delta n(x) \delta n(y) \right>$  integrated over all space
(this is trivially seen also in Eq.~(\ref{nnsolution})), which is a
consequence of charge conservation.

It is useful to rewrite the integral over $\tau$ in
Eq.~(\ref{Adefinition}) in terms of an integral over
temperature $T$, which is straightforward.  It is also useful to rewrite the
integral in the definition of $H$ in Eq.~(\ref{Hdef}) in the same
way. For that we need to know $dT/d\tau$. 
We shall neglect viscous corrections and use the ideal hydro
equation $\tau s= {\rm constant}$, or
\begin{eqnarray}
  \frac{d T}{d \tau} + \frac{T v_s^2(T)}{\tau} = 0,
\end{eqnarray}
where $v_s^2 \equiv \partial P / \partial \varepsilon$ is the speed of
sound (at zero charge density). Then,
\begin{multline}
H(T_\f ,T,k_\perp,k_\eta)\\=\int_{T_\f }^T \left(\frac{D(T') s(T')
  }{\tauf s_\f }\, k_\eta^2+ \frac{s_\f \tauf  D(T') }{s(T')}\,k_\perp^2\right)\frac{dT^\prime}{v_s^2(T^\prime) T^\prime}
\end{multline}
For convenience, we define dimensionless quantities $\hat{D}(T)\equiv D T$,  and $\hat{s}(T)\equiv  s(T)/T^3$.  Then $H(T_\f ,T,k_\perp,k_\eta)$ can be written as
\begin{multline}
H(T_\f ,T,k_\perp,k_\eta)= \frac{ k_\eta^2 }{\tauf s_\f }\int_{T_\f
}^T \frac{ \hat{D}(T') \hat{s}(T^\prime) T'}{v_s^2(T')}dT^\prime
\\+ s_\f \tauf k_\perp^2\int_{T_\f }^T \frac{\hat{D}(T')}{v_s^2(T'){ \hat{s}(T^\prime)(T^\prime)}^5 }dT^\prime,
\label{Hdimensionless}
\end{multline}
where the function $\hat{s}$ can be taken directly from lattice data
(see Section~\ref{LatticeEOSSec}). The speed of sound (at $ n=0$) can also be found from $\hat{s}$:
\begin{eqnarray}
v_s^2(T) =\left(3+\frac{d\ln\hat{s}}{d\ln T}\right)^{-1}.
\label{SoundSquared}
\end{eqnarray}
Combining Eq.(\ref{Adefinition}) and Eq.(\ref{Hdimensionless}), we find
\begin{widetext}
\begin{equation}
\tilde{A}(\tauf , k_\perp,k_\eta)=
\left[
 \frac{\chi_0T_0}{s_0} e^{-2H(T_\f ,T_0,k_\perp,k_\eta)}
-
\int_{T_\f }^{T_0}\!dT\frac{d}{dT}\left(\frac{\chi T}{s}\right)e^{-2H(T_\f ,T,k_\perp,k_\eta)}
\right].
\label{Afinal}
\end{equation}
\end{widetext}

\subsection{Discussion and Interpretation}
\label{sec:disc-interpr}
Eq.~(\ref{nnfinal}) along with the definitions
Eqs.~(\ref{Hdimensionless}),~(\ref{Afinal}) describes the charge
density correlation due to stochastic noise and is the main result of
this section.

The correlator in the square brackets in
Eq.~(\ref{nnfinal}) naturally
separates into a local part -- the first term, which is independent of $k$ and is thus a $\delta$-function in position space, and
the non-local part -- the second term, given by Eq.~(\ref{Afinal}), 
which vanishes at large $k$.

For a Boltzmann gas (which is a good approximation at freezeout) we can
identify the local term with the equilibrium self-correlation, which exists
even in a non-interacting gas.
On general grounds, the correlation function of a gas of particles in
equilibrium is expected to have such a delta-function term~\cite{LandauStatV5,LandauStatV9}
\begin{eqnarray}
  \left<\delta n(\vec{x}_1) \delta n(\vec{x}_2) \right> = \bar{n} \, \delta^{3}(\vec{x}_1 - \vec{x}_2) + ...
  \label{LandauCorrelation}
\end{eqnarray}
where $``..."$ denotes the correlations from interactions.  
This delta function is not due to a
correlation between {\em two} particles, as it is present even in a
free gas.  It is a trivial manifestation of statistical
\emph{fluctuations} in the gas, a reflection of the fact that
particles are trivially correlated with themselves (see \S116
of \cite{LandauStatV5}).  In a free Boltzmann gas
$\bar{n} = \chi T$ which is exactly the factor appearing in
Eq. (\ref{nnfinal}). The factor of $\tauf ^{-1}$ in
Eq. (\ref{nnfinal}) is the volume Jacobian factor, $1/\sqrt{-g}$,
for the delta-function in Bjorken coordinates.  Because
experimental measures count only {\em two}-particle correlations
it is necessary to separate the self-correlation term before comparing
with the data.  The separation of such a self-correlation term has
been also discussed in
\cite{Pratt:2012dz}, but not in \cite{Kapusta:2011gt,
  Kapusta:2012zb}.

We now turn to the non-local contribution to the correlator in
Eq. (\ref{Afinal}). We note that the dimensionless quantity $\chi T/s$
(charge susceptibility per entropy) and its $T$-dependence plays an
important role. 
The first term in $\tilde{A}$ is a three-dimensional negative Gaussian
with width (in position space) given by the diffusion distance over
the entire expansion history (since $\tau_0$), and a magnitude
controlled by the initial value of $\chi_0T_0/s_0$.  If $\chi T /s$
were constant, which would be the case in a conformal theory, and is
approximately the case in high temperature QGP, all non-trivial correlations resulting
from the diffusion history of hydrodynamical fluctuations would be
contained in this negative Gaussian.

The second term in Eq.~(\ref{Adefinition}) is due to the change of $\chi T /s$.
This term is a superposition of many Gaussians with
different widths and amplitudes.  Because this contribution
clearly requires the (charge-carrying) constituents of the plasma to
change, its main contribution comes from the QCD crossover region.  
We also find that this term 
gives the dominant contribution to the charge correlations in heavy-ion
collisions.  It is, therefore, essential for our calculation to know $\chi T /
s$ throughout the history of a heavy-ion collision, especially in the
crossover region, which is the subject of the next
section.

\subsection{The Temperature Dependence of Susceptibility per Entropy}
\label{LatticeEOSSec}

The behavior of entropy density $s$ and charge susceptibility $\chi$
as a function of temperature is easy to understand qualitatively and
semiquantitatively. In the crossover region the QCD matter undergoes a
smooth transition from the hadron gas to the QGP state. This leads to
a significant increase in the number of degrees of freedom (liberation
of color), i.e., growth of $s/T^3$. Although the number of charged
degrees of freedom also increases, their average charge is smaller in
QGP, and as a result the growth of $\chi/T^2$ is only moderate. The
growth of $s/T^3$ is much more significant (due to the gluons) and as
a result the dimensionless ratio $\chi T/s$ decreases with temperature
in the crossover region.

To make the above description more quantitative we can estimate $\chi
T/s$ in the QGP by considering ideal massless 
gases of gluons and quarks (3 massless flavors). Although this
approximation is only valid for asymptotically high $T$, it is
sufficient for our illustrative purposes. A straightforward
calculation leads to
\begin{equation}
  \label{eq:chisQGP}
  \mbox{QGP:}\quad   \frac{s}{T^{3}}= \frac{19\pi^2}{9};
\quad\frac{\chi}{T^{2}} = \frac23;
\quad \frac{\chi T}{s} = \frac{6}{19\pi^2}\,.
\end{equation}
For a rough estimate of these quantities in the hadron gas phase we
can take ideal gas of massless pions, for which we find
\begin{equation}
  \label{eq:chisHG}
  \mbox{Pion gas:}\quad  \frac{s}{T^{3}}= \frac{2\pi^2}{15};
\quad\frac{\chi}{T^{2}} = \frac13;
\quad \frac{\chi T}{s} = \frac{5}{2\pi^2}\,.
\end{equation}
We see that around 16-fold increase of $s/T^3$ in QGP relative to the
pion gas overwhelms the only 2-fold increase of $\chi/T^2$ leading to a
significant decrease of $\chi T/s$.

\begin{figure}
  \centering
\raisebox{.17\textwidth}{\bf(a)}
\includegraphics[width=0.32\textwidth]{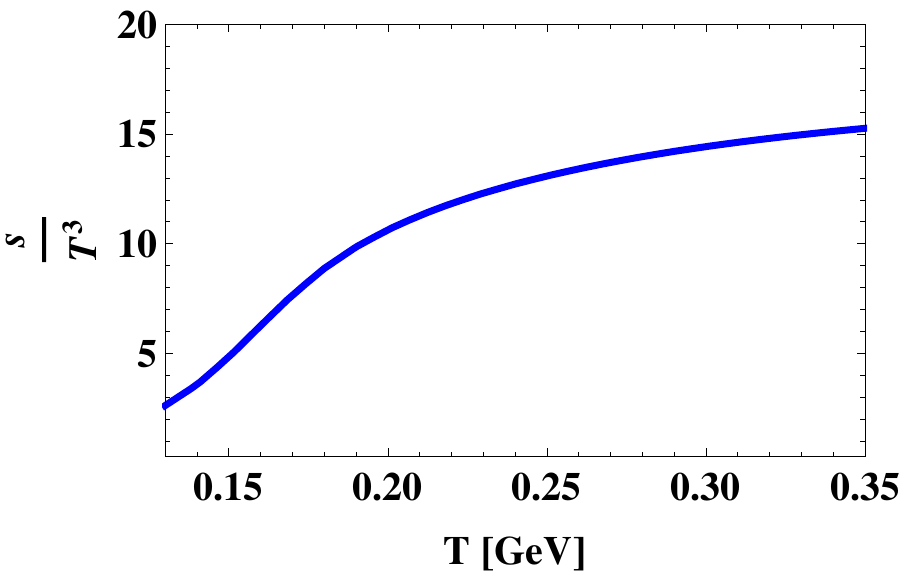}

\raisebox{.17\textwidth}{\bf(b)}
\includegraphics[width=0.32\textwidth]{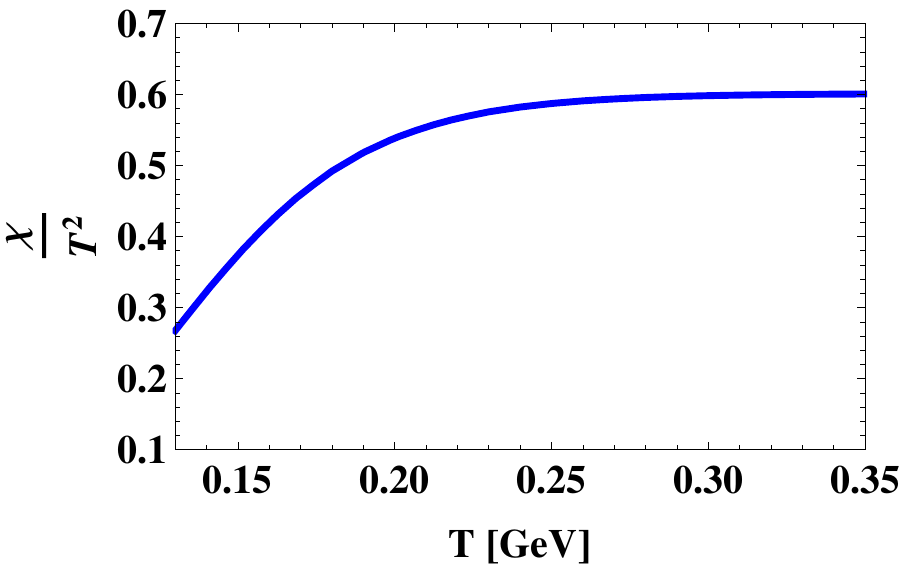}

\raisebox{.17\textwidth}{\bf(c)}
\includegraphics[width=0.32\textwidth]{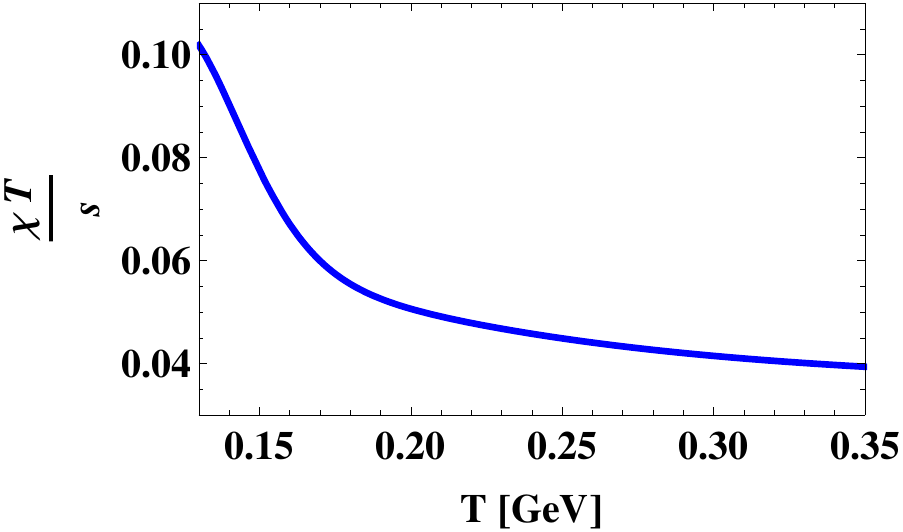}
\caption{(Color online).  The temperature dependence (in units of $T$)
  of entropy
  ($s/T^3$), charge susceptibility ($\chi/T^2$) taken from the lattice data
  \cite{Borsanyi:2010cj, Borsanyi:2011sw} and the resulting
  charge susceptibility per entropy (${\chi T}/{s}$) which
  we use in this paper.}
\label{ChiOverS}
\end{figure}

These simple estimates are in qualitative and semiquantitative
agreement with lattice QCD calculations
\cite{Cheng:2007jq,Cheng:2008zh,Bazavov:2009zn,Bazavov:2012jq,Borsanyi:2010cj}
which show that both entropy density and electric charge
susceptibility change significantly in the crossover
region. Figure~\ref{ChiOverS} shows our attempt to extract the ratio
$\chi T/s$ from these lattice results.  In our exploratory analysis we
shall ignore statistical or systematic errors on these data and use
equation of state shown in Fig.~\ref{ChiOverS} in our computations.

\section{Towards comparison with experiment}
\label{FreezeOutSec}
With the lattice data on electric susceptibility $\chi$, the entropy
density $s$ and the charge diffusion coefficient\footnote{Lattice data on the charge conductivity or the
diffusion coefficient $D$ is subject to more uncertainties due to
the analytic continuation from imaginary to real time and will be
discussed at the end of this section and in
Section~\ref{ConclusionSec}.} $D$, one can use the
results of Sec. \ref{BjorkenStochasticSec} to determine the spatial
correlations of the net charge due to hydrodynamic fluctuations.
These position space correlations need to be translated into particle
momentum space correlations which are measured experimentally in a
heavy-ion collision.  To achieve this goal we need to address several
important issues.

\subsection{Partial chemical equilibrium}
\label{sec:part-chem-equil}

The lattice equation of state, discussed in Sec.~\ref{LatticeEOSSec},
describes QCD matter in
full thermal and chemical equilibrium. Although this is a
reasonable approximation during much of the expansion history, it
breaks down after chemical freezeout. The rate of reactions
responsible for chemical equilibration (inelastic collisions)
is too slow in the hadronic phase to maintain chemical equilibrium. However,
the thermal (kinetic) equilibrium is maintained until later times.
In the intermediate region between the chemical and kinetic freezeout
the matter can be described using the so-called partial chemical
equilibrium (PCE) equation of state  \cite{Huovinen:2009yb}.

Rather than using the PCE equation of state we shall use a simpler
approach, based on the observation in Ref.~\cite{Teaney:2002aj} that
the PCE equation of state expressed as pressure vs energy density is very
similar to the full equilibrium (FE) equation of state. Since
hydrodynamic equations involve the equation of state $P(\varepsilon)$, their solution under PCE
should be similar to their solution under FE.~\footnote{To maintain
  simplicity and transparency of our results we make an assumption
  that the same is true for the charge susceptibility $\chi T$. The
  charge susceptibility was not discussed in
  Ref.~\cite{Teaney:2002aj}, and it would be interesting to test this
  natural assumption by applying the same methods to calculate $\chi
  T$. Deviations from this assumption can be included in our approach
  if necessary. } The difference is manifested when we ask what the
temperature is, given a value of the energy density, i.e., at a
given point in time in the expansion history.  Thus the actual kinetic
freezeout temperature $\Tkf$, which determines the final (observed)
momentum spectra of the particles, is different from the temperature,
$\Th$ which would correspond to the final energy density in full
equilibrium equation of state. The results of
Ref. \cite{Teaney:2002aj} suggest that for the kinetic freezeout
temperature $\Tkf\approx 100$ MeV the reasonable choice of the
corresponding temperature at which the FE equation of state gives the
same energy density is around $\Th\approx 130$ MeV. In this approach
one ends hydrodynamic evolution at a final temperature (hydrodynamic
freezeout) $\Th$ and implements the freezeout procedure with the
momentum spectra of particles determined by $\Tkf$
\cite{Teaney:2002aj}.

Additional conservation laws emerging under PCE are reflected in the
appearance of corresponding chemical potentials. For our study we will
need to use a chemical potential for the total pion number (the sum of the
numbers of $\pi^+$ and $\pi^-$), $\mu_\pi$. The value of this
chemical potential has also been estimated in
Ref.~\cite{Teaney:2002aj} in the range of $\mu_\pi\sim 80$ MeV. We shall
use the above values for $\Tkf$, $\Th$ and $\mu_\pi$ in our calculations
of the balance functions below.

\subsection{Transverse expansion}
\label{sec:transverse-expansion}

Since significant contribution to the balance function comes from the
crossover region, we must also take into account the fact that, due
to finite transverse size of the colliding nuclei, the expansion does not
remain purely Bjorken. Radial flow
becomes significant at times $\tau$ of order the initial radius of the
nucleus $R_0$ and the expansion approaches isotropic 3D
expansion at later times. As a result
the entropy density drops much faster, approximately as $\tau^{-3}$ 
during this later period, as opposed to $\tau^{-1}$ during
the 1D Bjorken period~\cite{Song:2007ux,Teaney:2009qa}.

Since analytical treatment of the full 3D expansion is beyond our
reach, we consider a simple idealized approximation by assuming that
the 3D stage of the expansion is short enough (due to fast drop in
entropy, and thus, temperature) that the diffusion during this period
can be neglected. We shall refer to this picture as the sudden transverse
expansion approximation.  In this idealized picture of the collision
the expansion follows the 1D Bjorken solution until a point in time which we
denote $\tau_{1D}$, upon which it undergoes sudden transverse expansion
and freezes out shortly thereafter at time $\tauf $ with a pattern of
flow given by the blast-wave ansatz.

We determine the charge correlator at time $\tau_{1D}$ (instead of
$\tauf$) using
Eq.(\ref{nnfinal}) and Eq.(\ref{Adefinition}) derived under conditions
of the $1D$ Bjorken expansion
\begin{multline}
\Big\langle\delta\tilde{ n}_n(\tau_{1D},k_\perp,k_\eta)\,\delta \tilde{n}^*_m(\tau_{1D},q_\perp,q_\eta)\Big\rangle =\hat{\delta}_{mn}(k,q) \\\times \frac{s_{1D}}{\tau_{1D}}\cdot \Bigg[\left(\frac{\chi T}{s}\right)_{1D}-\tilde{A}(\tau_{1D},k_\perp,k_\eta)\Bigg],
\label{nnfinal1D}
\end{multline}
where $\tilde{A}(\tau_{1D},k_\perp,k_\eta)$ is given by
Eq.~(\ref{Afinal}) with $\tauf$ replaced with $\tau_{1D}$. We then
should treat Eq.(\ref{nnfinal1D}) as the initial condition for the
period of the ``sudden'' 3D expansion. This fast expansion is
essentially adiabatic and thus the ratio $n/s$ is
conserved.  Since the entropy density drops from $s_{1D}$ to $s_\f $
during this period, the charge density must also drop by the same
factor. This means the charge correlator at time $\tauf $ obeys
\begin{eqnarray}
\Big\langle\delta\tilde{ n}_n\,\delta \tilde{n}^*_m\Big\rangle_\f =\frac{s^2_\f }{s^2_{1D}}\Big\langle\delta\tilde{ n}_n\,\delta\tilde{n}^*_m\Big\rangle_{1D}\,,
\end{eqnarray}
where $\left\langle\delta\tilde{
  n}_n\,\delta\tilde{n}^*_m\right\rangle_{1D}$ is given by
Eq.~(\ref{nnfinal1D}). Thus
\begin{multline}
\Big\langle\delta\tilde{ n}_n\,\delta \tilde{n}^*_m\Big\rangle_\f
=\hat{\delta}_{mn}(k,q) \cdot \frac{s_\f }{\tauf }\cdot\frac{s_\f \tauf
}{s_{1D}\tau_{1D}}\\\times
 \Bigg[\left(\frac{\chi T}{s}\right)_{1D}- \tilde{A}(\tau_{1D},k_\perp,k_\eta)\Bigg]\,.
\label{nnfinalFreeze-out1}
\end{multline}
The first (local in position space) term in Eq.~(\ref{nnfinalFreeze-out1})
contains the contribution of the self-correlation which we need to
subtract. As discussed in Section~\ref{sec:disc-interpr}, this self-correlation term is given by $\hat\delta_{mn}(k,q)(\chi
T/\tau)_\f $. Thus we write the charge density correlations at
freezeout as
\begin{eqnarray}
\Big\langle\delta\tilde{ n}_n\,\delta \tilde{n}^*_m\Big\rangle_\f
\equiv\hat{\delta}_{mn}(k,q)\left(\frac{\chi_\f  T_\f }{\tauf}
-\frac{s_\f }{\tauf }\tilde{A}_\f(k_\perp,k_\eta)\right),
\label{nnfinalFreeze-out2}
\end{eqnarray}
which defines two-particle hydrodynamic correlator $\tilde{A}_\f(k_\perp,k_\eta)$ at freezeout. Comparing Eq.(\ref{nnfinalFreeze-out1})
and Eq.(\ref{nnfinalFreeze-out2}), we find
\begin{multline}
\tilde{A}_\f(k_\perp,k_\eta)
=\frac{s_\f \tauf }{s_{1D}\tau_{1D}}\cdot \tilde{A}(\tau_{1D},k_\perp,k_\eta)
\\
+\left(\frac{\chi T}{s}\right)_\f 
-\frac{s_\f \tauf }{s_{1D}\tau_{1D}}\cdot\left(\frac{\chi
    T}{s}\right)_{1D}
\label{CPractical}
\end{multline}
and use it to calculate the balance function later in this section.

From Eq.(\ref{CPractical}) we see that the density-density
correlations built during the 1D Bjorken expansion period are diluted
due to the transverse expansion by a factor ${(s_\f \tauf)
}/{(s_{1D}\tau_{1D})}$ which would be equal to 1 if the system
continued pure 1D
expansion until freeze-out. Furthermore, the correlator $\tilde A_\f$
contains a local term, independent of $k$, because the last two terms
in Eq.~(\ref{CPractical}) do not cancel. This is
the contribution of the noise from the period of the sudden
transverse expansion. It is represented by a delta function in
position space because the noise is local and we neglected diffusion
during this short time, which would otherwise broaden the
delta function.

\subsection{Cooper-Frye freezeout}
In order to compare our results with experimental measurements, we
need to translate the hydrodynamic correlations in position space into
correlations in the kinematic (momentum) space of the observed particles.
For this purpose we
use the standard Cooper-Frye prescription for pions:
\begin{eqnarray}
\frac{dN_Q}{dy d\phi}=\frac{1}{(2\pi)^3}\int p_\perp dp_\perp \int d\sigma_\mu p^\mu\, f_Q(\vec{x},\vec{p})\,,
\label{CooperFrye}
\end{eqnarray}
where $f_Q= \exp\left\{ Q\mu/\Th+\mu_\pi/\Tkf- p_\mu u^\mu /\Tkf
\right\}$ is the equilibrium distribution function for pions carrying
charge $Q$ (equal to $\pm1$) in the Boltzmann
approximation~\footnote{The factors $1/T$ accompanying $\mu$ and
  $\mu_\pi$ in $f_Q$ reflect the definitions of these chemical
  potentials. While $\mu$ is defined in terms of the FE equation of state
used in hydrodynamics, the potential $\mu_\pi$ accounts for the pion
excess at kinetic freezeout due to PCE.}.

We have also defined kinematic rapidity as $y \equiv \tanh^{-1}\left(p^z/E\right) $,
kinematic azimuthal angle as $\phi\equiv
\tan^{-1}\left(p^y/p^x\right)$, and denoted the freezeout hypersurface
normal 4-vector as
$d\sigma_\mu$. The $p_\perp$
integration range is determined by experimental $p_\perp$ cuts.
We choose an isochronous freeze-out condition \footnote{
  For net charge correlations at zero chemical potential
  this is equivalent to isothermal
  freeze-out because fluctuations of temperature do not mix
  with charge fluctuations.} at $\tau = \tauf $, thus
\begin{equation}
   d\sigma_\mu p^\mu =\tauf  m_\perp d^2x_\perp d\eta \cosh(y-\eta),
\end{equation}
where $m_\perp \equiv \sqrt{p_\perp^2 + m^2}$, with
$m$ being the rest mass of the pion.

Since we are interested in the effect of the hydrodynamic
fluctuations, we expand the distribution function to linear order in
fluctuations of temperature, chemical potential, and fluid velocity.
If the average of the net chemical potential $\bar{\mu}$ is 0, then
only the chemical potential fluctuation survives in the difference
between particles and antiparticles:
\begin{multline}
\delta \frac{d\Nnet}{d  y d \phi}=\frac{2 \tauf }{(2\pi)^3 \Th
}\cdot\int m^2_\perp dm_\perp \int d^2x_\perp \int  d\eta  \\\times
\delta\mu(\tauf ,\vec x_\perp,\eta)\, \cosh(y-\eta) f_0(\vec{x},\vec{p}),
\label{3.3}
\end{multline}
where $f_0$ is the Boltzmann distribution function at $\mu=0$ and 
\begin{eqnarray}
  \Nnet \equiv N_+ - N_-.
\end{eqnarray}
Fluctuations of chemical potential are related to those of the charge
density by
$\delta n=\chi\,\delta \mu $.  

\subsection{Blast Wave}
\label{BlastWaveSec}

As we already discussed in Section~\ref{sec:transverse-expansion},
finiteness of the transverse size of the system leads to transverse expansion.
We shall describe the transverse flow velocity profile $v_r(r)$ using transverse
rapidity $\kappa(r)$
\begin{equation}\label{eq:kappa}
  v_r(r) = \frac{u^r}{u^\tau} \equiv \tanh \kappa(r).
\end{equation}
The distribution function $f_0$ can be then written as
\cite{Schnedermann:1993ws}
\begin{multline}
  f_0(\vec{x},\vec{p}) = \exp \big\{\hat\mu_\pi+\hat p_\perp
    \cos(\phi - \psi)\sinh \kappa_\f (r) \\- \hat m_\perp \cosh(y-\eta) \cosh \kappa_\f (r) \big\}
  \label{f0radialflow}
\end{multline}
where $\kappa_\f (r)$ describes the radial flow profile at kinetic freeze-out, 
 $\psi$ is  the position space
azimuthal angle characterizing the direction of the radius-vector
$\vec x$, and we introduced convenient dimensionless variables:
\begin{equation}
  \label{eq:hatm}
 \hat \mu_\pi= \mu_\pi/\Tkf
\quad  \hat m_\perp= m_\perp/\Tkf, 
\quad\hat p_\perp= p_\perp/\Tkf.
\end{equation}

We apply the standard blast-wave approach, i.e., we
specify the radial flow profile $\kappa(r)$ at freezeout by hand (as a
linear function of $r$) and limit the
transverse size of the system: $r<R$.
Such an approach is known to provide a reasonable approximation to
single particle observables computed using a full hydrodynamic
 solution which includes
transverse expansion \cite{Teaney:2002zt}.

Finally, we have
\begin{multline}\label{eq:dNnet}
\delta \frac{d\Nnet}{d  y d \phi}=\frac{ \tauf  \Tkf^3 R^2}{
  \chih  \Th } \int d^2\vec x_\perp \int_{-\infty}^{\infty}  d\eta
\\\times \delta n(\tauf ,\vec x_\perp,\eta)\, F(\vec{x},\vec{p}),
\end{multline}
where we introduced the function
\begin{multline}
  F(\vec{x},\vec{p}) \equiv \frac{1}{4 \pi^3 R^2}\int \hat
  m_\perp^2 d\hat m_\perp \cosh(y-\eta)\\\times f_0(\vec{x},\vec{p}) \Theta(R-r).
\label{Fdef}
\end{multline}
which acts as a kernel of the transformation from the position
variables $\vec x$ to kinematic variables $\vec p$. We normalized $F$
in such a way that its Fourier-Bessel transform is dimensionless (see below).

To proceed, we introduce Fourier-Bessel expansions for both $\delta n$
and $F$ in Eq.~(\ref{eq:dNnet}). Due to azimuthal and boost invariance (and integration over
$m_\perp$ in Eq.~(\ref{Fdef})) the function $F(\vec x, \vec p)$ depends only on three
arguments: $r$ and the differences $\phi - \psi$, and $y -
\eta$. We 
define the Fourier-Bessel transform with respect to these three
variables as $\tilde F_n(k_\perp,k_y)$ in terms of which we find,
substituting Eq.~(\ref{BesselTransform-f}) and using the closure
relation~(\ref{BesselClosure})
\begin{multline}\label{eq:dNet-Fn}
\delta \frac{d\Nnet}{d  y d \phi}=\frac{ \tauf  \Tkf^3 R^2}{ \chih \Th } \int_{-\infty}^{\infty} \frac{dk_y e^{i k_y y}}{2\pi}
\sum_{n} \frac{e^{ i n \phi}}{2\pi} \int_0^{\infty} k_\perp dk_\perp \\\times \delta \tilde{n}_n(\tauf ,k_\perp,k_y) \tilde{F}_n(k_\perp,k_{y}).
\end{multline}
For a given transverse flow profile at freezeout $\kappa_\f (r)$ in
Eq.~(\ref{eq:kappa}) we can obtain an expression for $\tilde{F}_n(k_\perp,k_{y})$ by
substituting Eq.~(\ref{f0radialflow}) into Eq.~(\ref{Fdef}) and
integrating over variables $(\phi-\psi)$ and $(y-\eta)$ in the
definition of the Fourier-Bessel transform Eq.~(\ref{BesselTransform})
\begin{multline}
  \tilde{F}_n(k_\perp,k_y) =-\frac{e^{\hat\mu_\pi}}{\pi^2}\int  {\hat m_\perp}^2
  d\hat m_\perp\int_0^1 \!\hat rd\hat r\, J_n(\hat k_\perp \hat
  r)\,\\\times\mathcal{I}_n(\hat p_\perp\sinh \kappa_\f )\, \mathcal{K}_{ik_y}'(\hat m_\perp\cosh\kappa_\f  ),
\end{multline}
where ${\cal I}$ is a modified Bessel function, ${\cal K}'$ is the
derivative of a modified Bessel function with respect to its argument and we
used convenient dimensionless variables defined in Eq.~(\ref{eq:hatm})
as well as
\begin{equation}
  \label{eq:hatk}
\quad \hat k_\perp = k_\perp R
\quad \mbox{and} \quad \hat r = r/R.
\end{equation}

It is also useful to note that the average value of the total number
of charged pions
\begin{equation}\label{eq:Nch}
 N_{\rm ch} \equiv N_+ + N_-
\end{equation}
 per unit
rapidity and azimuthal angle given by
\begin{multline}
\Big\langle\frac{dN_{\rm ch}}{dy d\phi}\Big\rangle=\frac{2 \tauf }{(2\pi)^3}\int  m_\perp^2 dm_\perp \int_{-\infty}^{\infty}  d\eta \int d^2x_\perp \\\times\cosh(y-\eta)f_0(\vec{x},\vec{p}) \Theta(R-r)
\end{multline}
can be also expressed as
\begin{equation}
  \label{eq:F00}
  \Big\langle\frac{dN_{\rm ch}}{dy d\phi}\Big\rangle = 
{\tauf  \Tkf^3 R^2}\tilde F_0(0,0)\,.
\end{equation}

\subsection{Particle Correlations and Balance Function}

Finally, to determine the particle correlations, we multiply two
fluctuations given by Eq.~(\ref{eq:dNet-Fn}), average over events and
express the correlator $\langle \delta \tilde n_n\delta\tilde
n_m\rangle$ using
Eq.~(\ref{nnfinal}), with the self-correlation
subtracted. 
The delta functions
in $\hat{\delta}_{mn}$ ensure that the result is only a
function of the rapidity difference $\Delta y \equiv y_2 - y_1$, and
angular difference $\Delta \phi = \phi_2 - \phi_1$ and we find
\begin{multline}\label{eq:dNnetdNnet}
\left\langle\delta\frac{d\Nnet}{dy_1 d\phi_1}\ \delta\frac{d\Nnet}{dy_2 d\phi_2} \right\rangle
=  -\left(\frac{\Tkf^3}{ \chih \Th }\right)^2  {s_\f  \tauf} R^2\\
\times\int_{-\infty}^{\infty} \frac{dk_y e^{i k_y \Delta y}}{2\pi}
\sum_{n} \frac{e^{ i n \Delta \phi}}{2\pi}
  \int_0^{\infty} \hat k_\perp d\hat k_\perp \\
\times \tilde{A}_\f(k_\perp,k_y) \tilde{F}_n(k_\perp,k_{y})\tilde{F}^*_n(k_\perp,k_{y})\,,
\end{multline}
where $k_\perp=\hat k_\perp/ R$ as in Eq.~(\ref{eq:hatk}).

When $\left<N_+ \right> = \left<N_- \right>$, the correlator
in Eq.~(\ref{eq:dNnetdNnet}) is related to
the balance function defined in
\cite{Bass:2000az,Jeon:2001ue} by
\begin{eqnarray}
  B( \Delta y,\Delta \phi) \equiv 
-\left\langle\delta\frac{d\Nnet}{dy_1 d\phi_1}\ \delta\frac{d\Nnet}{dy_2 d\phi_2} \right\rangle
  \left< \frac{d N_{\rm ch}}{dy d\phi} \right>^{-1}.
  \label{BalanceDefText}
\end{eqnarray}
This relationship is derived in Appendix \ref{BalanceAppendix}.
Finally, putting
Eqs.~(\ref{BalanceDefText}),~(\ref{eq:dNnetdNnet}) and~(\ref{eq:F00})
together, we find the expression for the balance function
\begin{multline}\label{BalanceFinal}
  B( \Delta y,\Delta \phi) = \frac{ \Tkf^3  s_\f }{ \chih^2 \Th^2 \tilde{F}_0(0,0)} \\
  \times \int_{-\infty}^{\infty} \frac{dk_y e^{i k_y \Delta y}}{2\pi}
\sum_{n} \frac{e^{ i n \Delta \phi}}{2\pi}
  \int_0^{\infty} \hat k_\perp d\hat k_\perp \\
  \times 
  \tilde{F}_n(k_\perp,k_{y})\tilde{F}^*_n(k_\perp,k_{y}) 
\tilde{A}_\f(k_\perp,k_y)\,.
\end{multline}
We use Eq. (\ref{BalanceFinal}) with $\tilde{A}_\f(k_\perp,k_y)$
given by Eq.~(\ref{CPractical}) to calculate the balance functions in
the next section.

We can calculate the rapidity and the azimuthal projections of the balance function
\begin{eqnarray}
B(\Delta y)&=&\int_{-\pi}^\pi d\Delta\phi \, B(\Delta y,\Delta \phi),
\notag
\\
B(\Delta\phi)&=&\int_{-\infty}^\infty d\Delta y\, B(\Delta y,\Delta \phi).
\label{BalanceFunctionProjection}
\end{eqnarray}
Integration over $\Delta \phi$ is equivalent to only
considering the $n=0$ moment in the summation of Eq.(\ref{BalanceFinal}),
while integration over $\Delta y$ is equivalent to setting $k_y=0$ instead
of performing an integral over $k_y$.

\subsection{Results}
\label{sec:results}

In order to illustrate the typical shape, width and magnitude of the balance
function arising due to the hydrodynamic fluctuations we calculate
this function using our semianalytical model of expansion described
above.  For central collisions at top RHIC energies, we set the time
when expansion stops being purely one-dimensional to $\tau_{1D}=7$ fm
and the corresponding temperature to $T_{1D}=150$ MeV. We set the
initial temperature to $T_0=350$ MeV. The hydro freezeout temperature
(see Section~\ref{sec:part-chem-equil}) is taken to be $\Th=130$ MeV
\cite{Teaney:2002aj}
We use the lattice data on entropy
density $s(T)$ \cite{Borsanyi:2010cj} and electric charge
susceptibility $\chi(T)$ \cite{Borsanyi:2011sw} as in
Fig.~\ref{ChiOverS}. We set the blast-wave
transverse flow profile to be linear $v_r=\frac{3}{2}\langle
\beta\rangle {r}/{R}$ with $\langle \beta\rangle=0.6$ and maximum
radius $R=12$~fm at $\tauf=12$~fm \cite{Shen:2012vn, Teaney:2002aj, Abelev:2008ab}.

\begin{figure*}[htpb]
\centering
\raisebox{2em}{\bf(a)}
\includegraphics[width=0.45\textwidth]{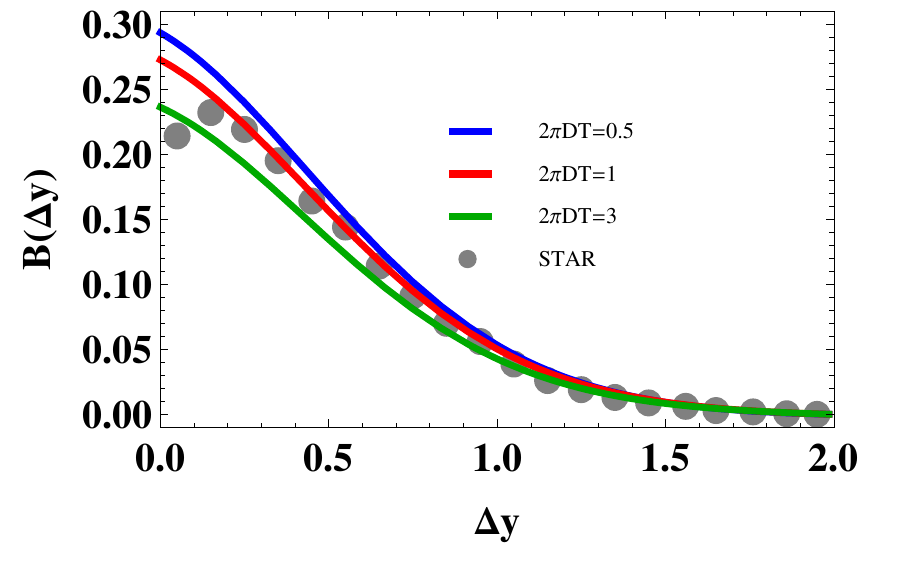}
\raisebox{2em}{\bf(b)}
\includegraphics[width=0.45\textwidth]{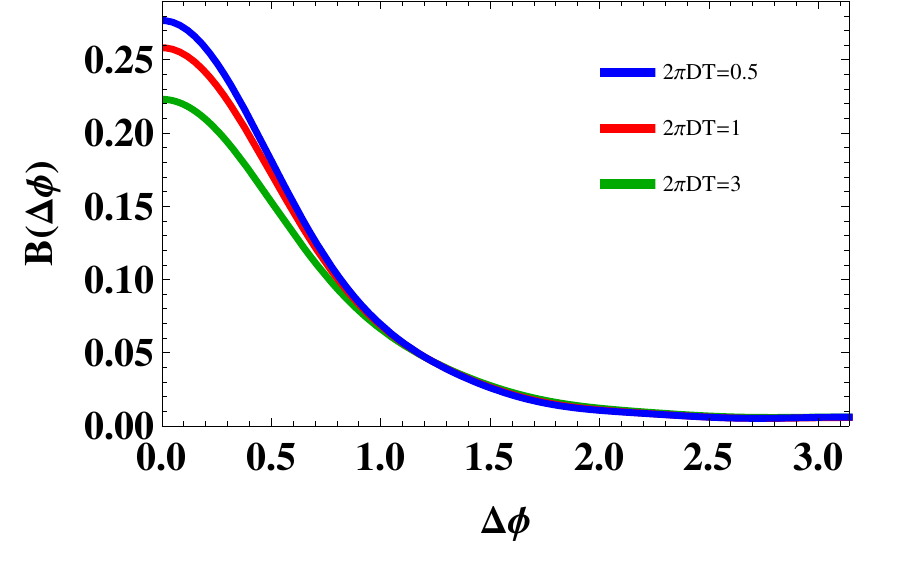}
\caption{(Color online).  (Left:) Rapidity projection of the pion
  balance function for different values of the diffusion coefficient.
  To compare with the experimental data from STAR
  \cite{Aggarwal:2010ya} we applied momentum cuts $0.2$ GeV$\leq
  p_\perp\leq 0.6$ GeV,  $|y|<1$ and efficiency
  correction 80\%. (Right:) azimuthal projection of the pion balance
  function for $0.2$ GeV$\leq p_\perp\leq 2$ GeV. }
\label{FinalBalanceFig}
\end{figure*}

In Fig.\ref{FinalBalanceFig}, we show the sensitivity of the balance
function to the charge diffusion coefficient, taking the dimensionless
combination $DT$ to be constant, with other parameters fixed.
In particular, we see that, for chosen values of parameters, the data
favors small values of the diffusion coefficient, $2\pi D T\sim 1$, which
is characteristic of a strongly coupled medium (short mean free
path).
Clearly, our semiquantitative analysis is not sufficient to pin down
the value of the diffusion coefficient with adequate precision, due to
the balance function's sensitivity to parameters which we fixed by hand (using typical
values obtained in numerical hydro simulations). However, our
results are indicative of the typical resolution one could achieve if
a more realistic numerical hydrodynamic simulation were to be
used instead of our simplified analytical model. We leave such quantitative
investigations to future work.

\section{Conclusions and discussion}
\label{ConclusionSec}

We showed that intrinsic hydrodynamic noise induces correlations
of charge fluctuations which are observable and  typically
quantified in terms of the charge balance functions. We have shown
how to calculate the noise contribution to the balance function and 
applied our method to a semianalytical model of hydrodynamic
expansion. The balance functions we obtain are in reasonable agreement
with experiments and our results suggest that a more realistic calculation may
allow one to determine or constrain the charge diffusion coefficient
$D$. Our semiquantitative analysis indicates that a small value of $D$,
characteristic of the strongly-coupled medium, is favored by the data.

Two main observations characterize the effect of the hydrodynamic
noise and diffusion on the charge balance functions. We find that the
magnitude of the balance function receives the most significant
contribution from the time interval during the expansion where the
charge susceptibility per entropy $\chi T/s$ changes most.   The
rapidity width of the balance function is determined by the diffusion
distance that the (originally local) correlation induced by noise
propagates during the time from its origin to the freezeout time.
\footnote{The
azimuthal width of the balance function is also sensitive to
diffusion, but is strongly affected by the radial flow. 
}

It is easy to understand that a change of the system's thermodynamic
state is necessary to produce a non-local correlation. Indeed, in a
static medium the correlations must be {\em local} (on hydrodynamic
scale) despite diffusion. This requires that the contributions from
successive time intervals cancel each other in a static medium,
leaving eventually only the (local) contribution from the most recent
time. We found that such cancellations could also occur in a medium
undergoing boost-invariant longitudinal expansion as long as $\chi
T\tau$ is constant (which is the same as $\chi T/s$ being constant up to
small viscous corrections).  In general, however, the expansion leads
to nonlocal correlations which carry the memory of the expansion.

One can think of this picture as the hydrodynamic description
of the mechanism of the suppression of charge fluctuations proposed
and analyzed in
Refs. \cite{Asakawa:2000wh,Jeon:2000wg,Shuryak:2000pd}. Indeed
the D-measure, $D_m$, introduced in Ref.\cite{Jeon:2000wg} is related
to the balance function as (see Appendix~\ref{BalanceAppendix})

\begin{eqnarray}
D_m\equiv 4\frac{\langle(\delta \Nnet)^2\rangle}{\langle N_{\rm ch}\rangle}
= 4 \left[1- \int_{-\infty}^{\infty} B(\Delta y) d\Delta y \right].
\label{DMeasureDef}
\end{eqnarray}
Therefore a positive balance function corresponds to suppression of net
charge fluctuations ($D_m<4$). The balance function provides 
differential phase-space information on the distribution of the anti-correlation
responsible for the suppression. Moreover, the
positivity of the balance function can be seen as a direct consequence
of the fact that $\chi T/s$ is smaller in QGP, i.e., $d(\chi
T/s)/dT<0$ (see Eq.~(\ref{Afinal})), which is the starting point of
the argument in \cite{Asakawa:2000wh,Jeon:2000wg}.

One can also view this hydrodynamic picture as effectively
representing the qualitative microscopic mechanism of charge balancing
described in
Refs.\cite{Bass:2000az,Jeon:2001ue,Pratt:2011bc,Pratt:2012dz}. The
advantage of hydrodynamic description is that it does not need to rely
on existence of quasiparticles. This is especially important because
both quark and hadron quasiparticle descriptions must break down
in the crossover region, and this is the region responsible for the major
contribution to the balance function. Our approach allows quantitative
description of these phenomena from first principles, i.e., from the
(lattice) equation of state and information on kinetic
coefficients, within a universal hydrodynamic formalism.

One of the many simplifying assumptions in our semianalytic
calculation has been the assumption that dimensionless combination
$DT$ is temperature independent. It is, perhaps, the easiest assumption
to relax, provided information of the temperature dependence of the
diffusion coefficient $D$ was available. Unlike the entropy and charge
susceptibility which, being static thermodynamic quantities, can be
reliably measured on the lattice, the diffusion coefficient is a
property of the real-time low-frequency response, which the Euclidean
time lattice calculation has well-known difficulties accessing. With
this caveat, it would be still interesting to extract the temperature
dependence of the charge diffusion coefficient from the existing
lattice data.

\begin{figure}
\raisebox{2em}{\bf(a)}
\includegraphics[width=0.32\textwidth]{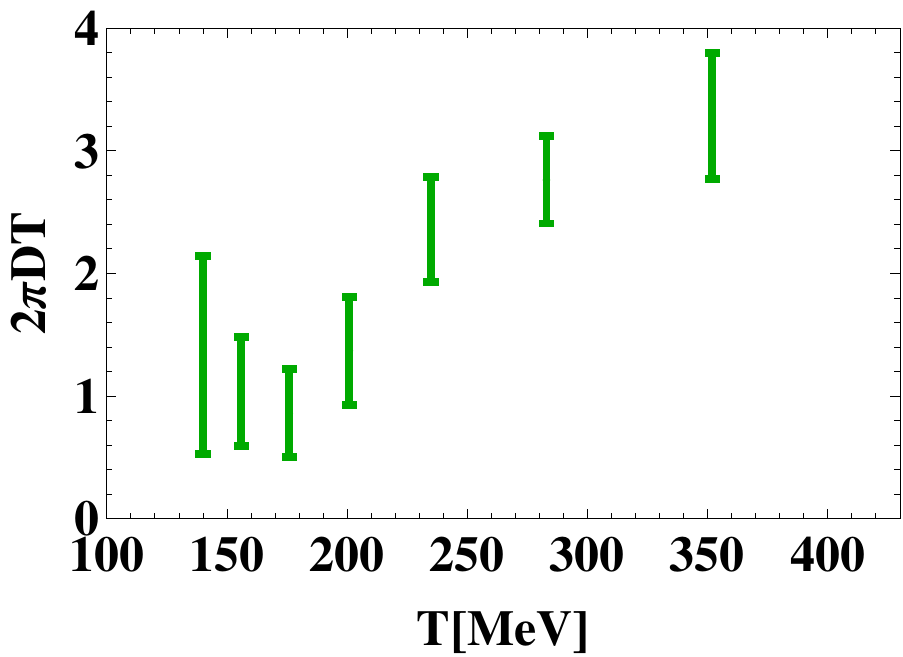}\\
\vskip 1em
\raisebox{2em}{\bf(b)}
\includegraphics[width=0.42\textwidth]{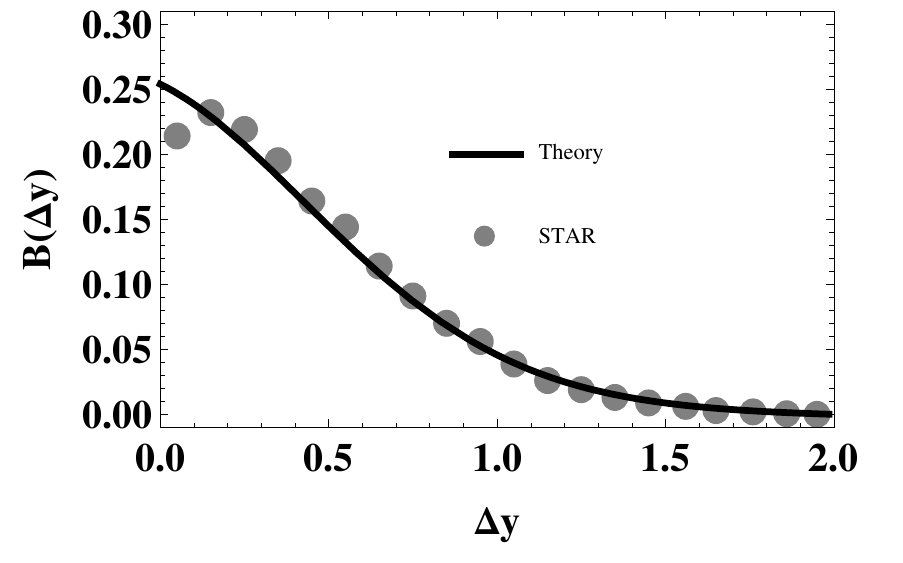}
\caption{(Color online.) (a) The temperature dependence of electric charge
  diffusion coefficient in units of $(2\pi T)^{-1}$ obtained by combining the lattice data from
  Refs.~\cite{Borsanyi:2011sw, Amato:2013naa}. (b) The balance
  function using the same parameters as for Fig.~\ref{FinalBalanceFig}
  but with the temperature dependent $2\pi D T$ taken from the lattice
  data. }
\label{DT}
\end{figure}

As a first exploratory step we attempted to combine recent lattice
data on electric conductivity \cite{Amato:2013naa} with the electric
susceptibility $\chi$ data from Ref.~\cite{Borsanyi:2011sw}. Using the
relation $D=\sigma/\chi$ we can then plot the temperature dependence
of the diffusion coefficient, or its dimensionless combination $2\pi DT$
(see Fig.\ref{DT}). 

Despite large error bars one can see that lattice
results suggest that the diffusion coefficient $D$ is indeed of order $1/2\pi
T$ in the crossover region, where we now know most of the
contribution to the balance function comes from. This is consistent
with the results of our comparison with experimental data in
Fig.~\ref{FinalBalanceFig}. 

Taking the lattice data as given (and ignoring the error bars) we can
also calculate the balance function using our semianalytic model. The
result plotted in Fig.~\ref{DT} shows a reasonable
agreement with the data.

An important improvement of our approach can be achieved by
implementing a more realistic pattern of radial flow, replacing the
blast-wave and sudden transverse expansion approximation. An approach 
based on the analytic solution proposed by Gubser {\it et al.}  
\cite{Gubser:2010ze, Gubser:2010ui} is tempting. However, the
limitation of this approach to a conformal equation of state is too
restrictive for our purpose, since the major contribution to the
balance function comes from the non-conformal (crossover) region.
A fully numerical hydrodynamic simulation with stochastic
noise will, of course, enable a quantitative comparison with
experiment. It would also allow extension of our results to
non-central (azimuthally asymmetric) collisions.

A natural application of the stochastic hydrodynamic approach is to
fluctuations near the QCD critical point \cite{Stephanov:2004wx} as
has been already initiated by \cite{Kapusta:2012zb}. The conductivity
$\sigma$ and susceptibility $\chi$ diverge at the critical point, leading
to the expected increase of the charge fluctuations at the critical
point. Such fluctuations are important signatures of the critical
point in heavy-ion collisions
\cite{Stephanov:1998dy,Stephanov:1999zu,Hatta:2003wn} and
understanding the effect of the time evolution
\cite{Berdnikov:1999ph,Stephanov:2009ra} and expansion on these
signatures is important for obtaining quantitative
predictions. However, such an application requires extension of the
formalism to non-zero baryon density where energy and flow velocity
fluctuations now mix with charge fluctuations \cite{Kapusta:2012zb}.

Another possible future direction is the examination of the balance
functions of kaons or protons \cite{Pratt:2012dz}, where the
fluctuation of strangeness or baryon number may become important.  It
should be straightforward to generalize our work to multiple conserved
charges.  We leave these topics for future studies.

\section*{Acknowledgments}
We thank D. Teaney, H.-U. Yee and Y. Yin for discussions.  
This work is supported by
the US Department of Energy under grant No.\ DE-FG0201ER41195.


  \bibliography{Bibliography/balancefunction}

\providecommand{\href}[2]{#2}\begingroup\raggedright\begin{thebibliography}{10}

\bibitem{Jeon:2003gk}
S.~Jeon and V.~Koch, ``{Event by event fluctuations},'' {\em in: Quark Gluon
  Plasma, edited by R. C. Hwa, X. N. Wang, Vol. 3 (World Scientific,
  Singapore)} (2004)  ,
\href{http://arxiv.org/abs/hep-ph/0304012}{{\tt arXiv:hep-ph/0304012
  [hep-ph]}}.

\bibitem{Jeon:2000wg}
S.~Jeon and V.~Koch, ``{Charged particle ratio fluctuation as a signal for
  QGP},'' \href{http://dx.doi.org/10.1103/PhysRevLett.85.2076}{{\em
  Phys.Rev.Lett.} {\bf 85} (2000)  2076--2079},
\href{http://arxiv.org/abs/hep-ph/0003168}{{\tt arXiv:hep-ph/0003168
  [hep-ph]}}.

\bibitem{Asakawa:2000wh}
M.~Asakawa, U.~W. Heinz, and B.~Muller, ``{Fluctuation probes of quark
  deconfinement},'' \href{http://dx.doi.org/10.1103/PhysRevLett.85.2072}{{\em
  Phys.Rev.Lett.} {\bf 85} (2000)  2072--2075},
\href{http://arxiv.org/abs/hep-ph/0003169}{{\tt arXiv:hep-ph/0003169
  [hep-ph]}}.

\bibitem{Abelev:2008jg}
{\bf STAR} Collaboration, B.~Abelev {\em et al.}, ``{Beam-Energy and
  System-Size Dependence of Dynamical Net Charge Fluctuations},''
  \href{http://dx.doi.org/10.1103/PhysRevC.79.024906}{{\em Phys.Rev.} {\bf C79}
  (2009)  024906},
\href{http://arxiv.org/abs/0807.3269}{{\tt arXiv:0807.3269 [nucl-ex]}}.

\bibitem{Abelev:2012pv}
{\bf ALICE Collaboration} Collaboration, B.~Abelev {\em et al.}, ``{Net-Charge
  Fluctuations in Pb-Pb collisions at $\sqrt{s}_{NN} = 2.76$ TeV},''
  \href{http://dx.doi.org/10.1103/PhysRevLett.110.152301}{{\em Phys.Rev.Lett.}
  {\bf 110} (2013) no.~15, 152301},
\href{http://arxiv.org/abs/1207.6068}{{\tt arXiv:1207.6068 [nucl-ex]}}.

\bibitem{Bass:2000az}
S.~A. Bass, P.~Danielewicz, and S.~Pratt, ``{Clocking hadronization in
  relativistic heavy ion collisions with balance functions},''
  \href{http://dx.doi.org/10.1103/PhysRevLett.85.2689}{{\em Phys.Rev.Lett.}
  {\bf 85} (2000)  2689--2692},
\href{http://arxiv.org/abs/nucl-th/0005044}{{\tt arXiv:nucl-th/0005044
  [nucl-th]}}.

\bibitem{Jeon:2001ue}
S.~Jeon and S.~Pratt, ``{Balance functions, correlations, charge fluctuations
  and interferometry},''
  \href{http://dx.doi.org/10.1103/PhysRevC.65.044902}{{\em Phys.Rev.} {\bf C65}
  (2002)  044902},
\href{http://arxiv.org/abs/hep-ph/0110043}{{\tt arXiv:hep-ph/0110043
  [hep-ph]}}.

\bibitem{Bozek:2012en}
P.~Bozek and W.~Broniowski, ``{Charge conservation and the shape of the ridge
  of two-particle correlations in relativistic heavy-ion collisions},''
  \href{http://dx.doi.org/10.1103/PhysRevLett.109.062301}{{\em Phys.Rev.Lett.}
  {\bf 109} (2012)  062301},
\href{http://arxiv.org/abs/1204.3580}{{\tt arXiv:1204.3580 [nucl-th]}}.

\bibitem{Pratt:2012dz}
S.~Pratt, ``{Identifying the Charge Carriers of the Quark-Gluon Plasma},''
  \href{http://dx.doi.org/10.1103/PhysRevLett.108.212301}{{\em Phys.Rev.Lett.}
  {\bf 108} (2012)  212301},
\href{http://arxiv.org/abs/1203.4578}{{\tt arXiv:1203.4578 [nucl-th]}}.

\bibitem{Abelev:2009jv}
{\bf STAR Collaboration} Collaboration, B.~Abelev {\em et al.},
  ``{Three-particle coincidence of the long range pseudorapidity correlation in
  high energy nucleus-nucleus collisions},''
  \href{http://dx.doi.org/10.1103/PhysRevLett.105.022301}{{\em Phys.Rev.Lett.}
  {\bf 105} (2010)  022301},
\href{http://arxiv.org/abs/0912.3977}{{\tt arXiv:0912.3977 [hep-ex]}}.

\bibitem{Timmins:2011um}
{\bf ALICE} Collaboration, A.~R. Timmins, ``{Untriggered di-hadron correlations
  in Pb-Pb collisions at $\sqrt{s_{NN}} =$ 2.76 TeV from ALICE},'' {\em J.
  Phys. G: Nucl. Part. Phys.} {\bf 38} (2011)  124093,
\href{http://arxiv.org/abs/1106.6057}{{\tt arXiv:1106.6057 [nucl-ex]}}.

\bibitem{Aggarwal:2010ya}
{\bf STAR} Collaboration, M.~Aggarwal {\em et al.}, ``{Balance Functions from
  Au$+$Au, $d+$Au, and $p+p$ Collisions at $\sqrt{s_{NN}}$ = 200 GeV},''
  \href{http://dx.doi.org/10.1103/PhysRevC.82.024905}{{\em Phys.Rev.} {\bf C82}
  (2010)  024905},
\href{http://arxiv.org/abs/1005.2307}{{\tt arXiv:1005.2307 [nucl-ex]}}.

\bibitem{Abelev:2013csa}
{\bf ALICE Collaboration} Collaboration, B.~Abelev {\em et al.}, ``{Charge
  correlations using the balance function in Pb-Pb collisions at
  $\sqrt{s_{NN}}$ = 2.76 TeV},''
  \href{http://dx.doi.org/10.1016/j.physletb.2013.05.039}{{\em Phys.Lett.} {\bf
  B723} (2013)  267--279},
\href{http://arxiv.org/abs/1301.3756}{{\tt arXiv:1301.3756 [nucl-ex]}}.

\bibitem{Kapusta:2011gt}
J.~Kapusta, B.~Muller, and M.~Stephanov, ``{Relativistic Theory of Hydrodynamic
  Fluctuations with Applications to Heavy Ion Collisions},''
  \href{http://dx.doi.org/10.1103/PhysRevC.85.054906}{{\em Phys.Rev.} {\bf C85}
  (2012)  054906},
\href{http://arxiv.org/abs/1112.6405}{{\tt arXiv:1112.6405 [nucl-th]}}.

\bibitem{LandauStatV9}
E.~M. {Lifshitz} and L.~P. {Pitaevskii}, {\em Statistical physics: part 2,
  volume 9}.
\newblock Course of theoretical physics, Oxford: Pergamon Press, 1980.

\bibitem{Bjorken:1982qr}
J.~Bjorken, ``{Highly Relativistic Nucleus-Nucleus Collisions: The Central
  Rapidity Region},''
\href{http://dx.doi.org/10.1103/PhysRevD.27.140}{{\em Phys.Rev.} {\bf D27}
  (1983)  140--151}.

\bibitem{Borsanyi:2010cj}
S.~Borsanyi, G.~Endrodi, Z.~Fodor, A.~Jakovac, S.~D. Katz, {\em et al.}, ``{The
  QCD equation of state with dynamical quarks},''
  \href{http://dx.doi.org/10.1007/JHEP11(2010)077}{{\em JHEP} {\bf 11} (2010)
  077},
\href{http://arxiv.org/abs/1007.2580}{{\tt arXiv:1007.2580 [hep-lat]}}.

\bibitem{Borsanyi:2011sw}
S.~Borsanyi, Z.~Fodor, S.~D. Katz, S.~Krieg, C.~Ratti, {\em et al.},
  ``{Fluctuations of conserved charges at finite temperature from lattice
  QCD},'' \href{http://dx.doi.org/10.1007/JHEP01(2012)138}{{\em JHEP} {\bf 01}
  (2012)  138},
\href{http://arxiv.org/abs/1112.4416}{{\tt arXiv:1112.4416 [hep-lat]}}.

\bibitem{LandauStatV5}
L.~D. Landau and E.~M. Lifshitz, {\em Statistical physics: part 1, volume 5}.
\newblock Course of theoretical physics, Oxford: Pergamon Press, 1980.

\bibitem{GardinerStochastic}
C.~{Gardiner}, {\em {Stochastic Methods}}.
\newblock Springer; 4th ed., 2009.

\bibitem{Fox:1978}
R.~F. Fox, ``Gaussian stochastic processes in physics,'' {\em Phys. Rep.} {\bf
  48} (1978)  179--283.

\bibitem{Kapusta:2012zb}
J.~I. Kapusta and J.~M. Torres-Rincon, ``{Thermal Conductivity and Chiral
  Critical Point in Heavy Ion Collisions},''
  \href{http://dx.doi.org/10.1103/PhysRevC.86.054911}{{\em Phys.Rev.} {\bf C86}
  (2012)  054911},
\href{http://arxiv.org/abs/1209.0675}{{\tt arXiv:1209.0675 [nucl-th]}}.

\bibitem{Cheng:2007jq}
M.~Cheng, N.~Christ, S.~Datta, J.~van~der Heide, C.~Jung, {\em et al.}, ``{The
  QCD equation of state with almost physical quark masses},''
  \href{http://dx.doi.org/10.1103/PhysRevD.77.014511}{{\em Phys.Rev.} {\bf D77}
  (2008)  014511},
\href{http://arxiv.org/abs/0710.0354}{{\tt arXiv:0710.0354 [hep-lat]}}.

\bibitem{Cheng:2008zh}
M.~Cheng, P.~Hendge, C.~Jung, F.~Karsch, O.~Kaczmarek, {\em et al.}, ``{Baryon
  Number, Strangeness and Electric Charge Fluctuations in QCD at High
  Temperature},'' \href{http://dx.doi.org/10.1103/PhysRevD.79.074505}{{\em
  Phys.Rev.} {\bf D79} (2009)  074505},
\href{http://arxiv.org/abs/0811.1006}{{\tt arXiv:0811.1006 [hep-lat]}}.

\bibitem{Bazavov:2009zn}
A.~Bazavov, T.~Bhattacharya, M.~Cheng, N.~Christ, C.~DeTar, {\em et al.},
  ``{Equation of state and QCD transition at finite temperature},''
  \href{http://dx.doi.org/10.1103/PhysRevD.80.014504}{{\em Phys.Rev.} {\bf D80}
  (2009)  014504},
\href{http://arxiv.org/abs/0903.4379}{{\tt arXiv:0903.4379 [hep-lat]}}.

\bibitem{Bazavov:2012jq}
{\bf HotQCD} Collaboration, A.~Bazavov {\em et al.}, ``{Fluctuations and
  Correlations of net baryon number, electric charge, and strangeness: A
  comparison of lattice QCD results with the hadron resonance gas model},''
  \href{http://dx.doi.org/10.1103/PhysRevD.86.034509}{{\em Phys.Rev.} {\bf D86}
  (2012)  034509},
\href{http://arxiv.org/abs/1203.0784}{{\tt arXiv:1203.0784 [hep-lat]}}.

\bibitem{Huovinen:2009yb}
P.~Huovinen and P.~Petreczky, ``{QCD Equation of State and Hadron Resonance
  Gas},'' \href{http://dx.doi.org/10.1016/j.nuclphysa.2010.02.015}{{\em
  Nucl.Phys.} {\bf A837} (2010)  26--53},
\href{http://arxiv.org/abs/0912.2541}{{\tt arXiv:0912.2541 [hep-ph]}}.

\bibitem{Teaney:2002aj}
D.~Teaney, ``{Chemical freezeout in heavy ion collisions},''
\href{http://arxiv.org/abs/nucl-th/0204023}{{\tt arXiv:nucl-th/0204023
  [nucl-th]}}.

\bibitem{Song:2007ux}
H.~Song and U.~W. Heinz, ``{Causal viscous hydrodynamics in 2+1 dimensions for
  relativistic heavy-ion collisions},''
  \href{http://dx.doi.org/10.1103/PhysRevC.77.064901}{{\em Phys.Rev.} {\bf C77}
  (2008)  064901},
\href{http://arxiv.org/abs/0712.3715}{{\tt arXiv:0712.3715 [nucl-th]}}.

\bibitem{Teaney:2009qa}
D.~A. Teaney, ``{Viscous Hydrodynamics and the Quark Gluon Plasma},'' {\em in:
  Quark Gluon Plasma, edited by R.C. Hwa, X.N. Wang, Vol. 4, (World Scientific,
  Singapore)} (2010)  ,
\href{http://arxiv.org/abs/0905.2433}{{\tt arXiv:0905.2433 [nucl-th]}}.

\bibitem{Schnedermann:1993ws}
E.~Schnedermann, J.~Sollfrank, and U.~W. Heinz, ``{Thermal phenomenology of
  hadrons from 200-A/GeV S+S collisions},''
  \href{http://dx.doi.org/10.1103/PhysRevC.48.2462}{{\em Phys.Rev.} {\bf C48}
  (1993)  2462--2475},
\href{http://arxiv.org/abs/nucl-th/9307020}{{\tt arXiv:nucl-th/9307020
  [nucl-th]}}.

\bibitem{Teaney:2002zt}
D.~Teaney, ``{Viscous corrections to spectra, elliptic flow, and HBT radii},''
  \href{http://dx.doi.org/10.1016/S0375-9474(02)01502-6}{{\em Nucl.Phys.} {\bf
  A715} (2003)  817c--820c},
\href{http://arxiv.org/abs/nucl-th/0209024}{{\tt arXiv:nucl-th/0209024
  [nucl-th]}}.

\bibitem{Shen:2012vn}
C.~Shen and U.~Heinz, ``{Collision Energy Dependence of Viscous Hydrodynamic
  Flow in Relativistic Heavy-Ion Collisions},''
  \href{http://dx.doi.org/10.1103/PhysRevC.86.049903,
  10.1103/PhysRevC.85.054902}{{\em Phys.Rev.} {\bf C85} (2012)  054902},
\href{http://arxiv.org/abs/1202.6620}{{\tt arXiv:1202.6620 [nucl-th]}}.

\bibitem{Abelev:2008ab}
{\bf STAR} Collaboration, B.~Abelev {\em et al.}, ``{Systematic Measurements of
  Identified Particle Spectra in $p p, d^+$ Au and Au+Au Collisions from
  STAR},'' \href{http://dx.doi.org/10.1103/PhysRevC.79.034909}{{\em Phys.Rev.}
  {\bf C79} (2009)  034909},
\href{http://arxiv.org/abs/0808.2041}{{\tt arXiv:0808.2041 [nucl-ex]}}.

\bibitem{Shuryak:2000pd}
E.~V. Shuryak and M.~A. Stephanov, ``{When can long range charge fluctuations
  serve as a QGP signal?},''
  \href{http://dx.doi.org/10.1103/PhysRevC.63.064903}{{\em Phys.Rev.} {\bf C63}
  (2001)  064903},
\href{http://arxiv.org/abs/hep-ph/0010100}{{\tt arXiv:hep-ph/0010100
  [hep-ph]}}.

\bibitem{Pratt:2011bc}
S.~Pratt, ``{General Charge Balance Functions, A Tool for Studying the Chemical
  Evolution of the Quark-Gluon Plasma},''
  \href{http://dx.doi.org/10.1103/PhysRevC.85.014904}{{\em Phys.Rev.} {\bf C85}
  (2012)  014904},
\href{http://arxiv.org/abs/1109.3647}{{\tt arXiv:1109.3647 [nucl-th]}}.

\bibitem{Amato:2013naa}
A.~Amato, G.~Aarts, C.~Allton, P.~Giudice, S.~Hands, {\em et al.},
  ``{Electrical conductivity of the quark-gluon plasma across the deconfinement
  transition},'' \href{http://dx.doi.org/10.1103/PhysRevLett.111.172001}{{\em
  Phys.Rev.Lett.} {\bf 111} (2013)  172001},
\href{http://arxiv.org/abs/1307.6763}{{\tt arXiv:1307.6763 [hep-lat]}}.

\bibitem{Gubser:2010ze}
S.~S. Gubser, ``{Symmetry constraints on generalizations of Bjorken flow},''
  \href{http://dx.doi.org/10.1103/PhysRevD.82.085027}{{\em Phys.Rev.} {\bf D82}
  (2010)  085027},
\href{http://arxiv.org/abs/1006.0006}{{\tt arXiv:1006.0006 [hep-th]}}.

\bibitem{Gubser:2010ui}
S.~S. Gubser and A.~Yarom, ``{Conformal hydrodynamics in Minkowski and de
  Sitter spacetimes},''
  \href{http://dx.doi.org/10.1016/j.nuclphysb.2011.01.012}{{\em Nucl.Phys.}
  {\bf B846} (2011)  469--511},
\href{http://arxiv.org/abs/1012.1314}{{\tt arXiv:1012.1314 [hep-th]}}.

\bibitem{Stephanov:2004wx}
M.~A. Stephanov, ``{QCD phase diagram and the critical point},''
  \href{http://dx.doi.org/10.1142/S0217751X05027965}{{\em Prog. Theor. Phys.
  Suppl.} {\bf 153} (2004)  139--156},
\href{http://arxiv.org/abs/hep-ph/0402115}{{\tt arXiv:hep-ph/0402115}}.

\bibitem{Stephanov:1998dy}
M.~A. Stephanov, K.~Rajagopal, and E.~V. Shuryak, ``{Signatures of the
  tricritical point in {QCD}},''
  \href{http://dx.doi.org/10.1103/PhysRevLett.81.4816}{{\em Phys. Rev. Lett.}
  {\bf 81} (1998)  4816--4819},
\href{http://arxiv.org/abs/hep-ph/9806219}{{\tt arXiv:hep-ph/9806219}}.

\bibitem{Stephanov:1999zu}
M.~A. Stephanov, K.~Rajagopal, and E.~V. Shuryak, ``{Event-by-event
  fluctuations in heavy ion collisions and the {QCD} critical point},''
  \href{http://dx.doi.org/10.1103/PhysRevD.60.114028}{{\em Phys. Rev.} {\bf
  D60} (1999)  114028},
\href{http://arxiv.org/abs/hep-ph/9903292}{{\tt arXiv:hep-ph/9903292}}.

\bibitem{Hatta:2003wn}
Y.~Hatta and M.~A. Stephanov, ``{Proton number fluctuation as a signal of the
  QCD critical end-point},''
  \href{http://dx.doi.org/10.1103/PhysRevLett.91.102003}{{\em Phys. Rev. Lett.}
  {\bf 91} (2003)  102003},
\href{http://arxiv.org/abs/hep-ph/0302002}{{\tt arXiv:hep-ph/0302002}}.

\bibitem{Berdnikov:1999ph}
B.~Berdnikov and K.~Rajagopal, ``{Slowing out of equilibrium near the QCD
  critical point},'' \href{http://dx.doi.org/10.1103/PhysRevD.61.105017}{{\em
  Phys. Rev.} {\bf D61} (2000)  105017},
\href{http://arxiv.org/abs/hep-ph/9912274}{{\tt arXiv:hep-ph/9912274}}.

\bibitem{Stephanov:2009ra}
M.~Stephanov, ``{Evolution of fluctuations near QCD critical point},''
  \href{http://dx.doi.org/10.1103/PhysRevD.81.054012}{{\em Phys.Rev.} {\bf D81}
  (2010)  054012},
\href{http://arxiv.org/abs/0911.1772}{{\tt arXiv:0911.1772 [hep-ph]}}.

\end{thebibliography}\endgroup

  \begin{widetext}

  \end{widetext}

\appendix

\section{A review of balance functions}
\label{BalanceAppendix}

For completeness we review here the definitions and properties of the
balance functions~\cite{Bass:2000az,Jeon:2001ue}.

To define the balance function
 we divide the phase space occupied by
particles produced in a heavy-ion collision into (infinitesimally)
small cells. For the purpose of this paper we consider cells in
rapidity $y$ and azimuthal angle $\phi$ (but integrated over
transverse momentum) and denote the coordinates of the cell
$\Gamma=(y, \phi)$ and the volume of the cell $d\Gamma=dy\, d\phi$.  We
denote the number of particles of charge $a=+,-$ in a cell as
$dN^a(\Gamma)$ and its event average $\langle
dN^a(\Gamma)\rangle$. Since $\langle dN^a(\Gamma)\rangle\sim
d\Gamma$ is infinitesimally small, the probability of finding more than one particle in a
cell is negligible (${\cal O}(d\Gamma^2)$ for two or more particles) and the average $\langle dN^a(\Gamma)\rangle\ll1$
is also the probability to find a particle of charge $a$ in the
cell. 

The conditional probability of finding a particle of charge $b$
in another cell $\Gamma_2$ given a particle of charge $a$ in the cell
$\Gamma_1$ can be found as $\langle
dN^b(\Gamma_2)\,dN^a(\Gamma_1)\rangle/\langle dN^a(\Gamma_1)\rangle$
which is easy to understand keeping in mind that $dN^a(\Gamma)$ is
 either $0$ or (rarely)~$1$. The balance function
defined on a pair of cells is given by:
\begin{multline}
  \label{eq:B12}
  B(\Gamma_2,\Gamma_1)=\frac1{2}\sum_{a=+,-}
\frac{\langle dN^{-a}_2 dN^a_1\rangle-\langle dN^a_2 dN^a_1\rangle}
{d\Gamma_2\langle dN^a_1\rangle}\\
=
\frac{\langle n^{-a}_2 n^a_1\rangle-\langle n^a_2 n^a_1\rangle}
{\langle n^a_1\rangle}
\end{multline}
where we used a shorthand $dN^a_i\equiv dN^a(\Gamma_i)$ and introduced
density per phase space volume $n\equiv dN/d\Gamma$.  The balance
function measures a difference in conditional probabilities of finding
a particle of the opposite charge $-a$ vs the same charge $a$ in
the cell $\Gamma_2$ given a particle of the charge $a$ in cell
$\Gamma_1$. This probability is proportional to the volume $d\Gamma_2$
of the cell and is infinitesimally small, while its ratio to
$d\Gamma_2$, as in Eq.~(\ref{eq:B12}), is finite.

Since we are considering a case when $\mu=0$, we can use $\langle
n^-_1\rangle=\langle
n^+_1\rangle$ 
to simplify Eq.~(\ref{eq:B12}):
\begin{equation}
  \label{eq:B12-simple}
  B(\Gamma_2,\Gamma_1)=
-\frac
{\left\langle (n^{+}_2 - n^-_2)(n^{+}_1 - n^-_1)\right\rangle}
{2\langle n^+_1\rangle}
= -\frac
{\left\langle n^{\rm net}_2\,n^{\rm net}_1\right\rangle}
{\langle n^{\rm ch}_1\rangle},
\end{equation}
Since $n^{\rm net}=d\Nnet/dyd\phi$ and $\Nnet=\delta \Nnet$
($\langle\Nnet\rangle=0$), this gives us
equation~(\ref{BalanceDefText}) used in the text.

One also defines the balance function as a function of the phase space
displacement $\Delta\Gamma\equiv
\Gamma_2-\Gamma_1=(y_2-y_1,\phi_2-\phi_1)$ by summing in
Eq.~(\ref{eq:B12}) over all cells $\Gamma_1$ and $\Gamma_2$ separated
by $\Delta \Gamma$. To obtain a finite result for infinitely many
infinitesimally small cells ($d\Gamma_i\to0$) we multiply by
$d\Gamma_1d\Gamma_2$. We can then write this summation as an
integral:
\begin{equation}
  \label{eq:B-Delta}
  B(\Delta\Gamma)=\frac1{\int d\Gamma}\,
\int \!d\Gamma_2\int \!d\Gamma_1\
  \delta(\Gamma_2-\Gamma_1-\Delta\Gamma)\,B(\Gamma_2,\Gamma_1)\,
.
\end{equation}
The normalization factor
is chosen in such a way that the result tends to a finite limit with
increasing total phase-space volume ($\int d\Gamma$).

The expression in Eq.~(\ref{eq:B-Delta})
 simplifies in the case of azimuthal and boost
($\Gamma\to\Gamma+\Delta \Gamma$) invariance. 
Since in this case the balance function $B(\Gamma_2,\Gamma_1)$ can
only depend on $\Delta\Gamma$ we find from Eq.~(\ref{eq:B-Delta}), simply,
 \begin{equation}
   \label{eq:B-Delta2}
   B(\Delta\Gamma)=B(\Gamma+\Delta \Gamma,\Gamma),
 \end{equation}
for any $\Gamma$.

The derivation above assumes that the rapidity acceptance window is infinite:
$y\in(-\infty,\infty)$, or more precisely, is much larger than the
rapidity range of the balance function $B(\Delta y,\Delta\phi)$. In
practice, the rapidity interval has a finite width $Y$. Still assuming boost
invariance, but integrating in Eq.~(\ref{eq:B-Delta}) over the finite
rapidity window of width $Y$ we find the balance function in a finite rapidity
acceptance:
\begin{equation}
  \label{eq:BY}
  B(\Delta\Gamma)_Y = B(\Delta\Gamma)_\infty\,\frac{Y-\Delta y}{Y}\,
\end{equation}
where we used $\int dy= Y$ and $\int dy_2\int dy_1\
  \delta(y_2-y_1-\Delta y)=Y-\Delta y$.

To express the D-measure $D_m$ \cite{Jeon:2000wg} in terms of the
balance function we substitute $\Nnet=\int\! d\Gamma\, n^{\rm
  net}(\Gamma)$ and $N_{\rm ch}=\int\! d\Gamma\, n^{\rm
  ch}(\Gamma)$ into the definition
\begin{equation}
  \label{eq:Dm}
  D_m \equiv 4\frac{\langle(\delta \Nnet)^2\rangle}{\langle N_{\rm
      ch}\rangle}
=  \frac4{\int d\Gamma}\,
\int \!d\Gamma_2\int \!d\Gamma_1\
\frac{\left\langle n^{\rm net}_2\,n^{\rm net}_1\right\rangle}
{\langle n^{\rm ch}\rangle}
\end{equation}
The integrand is $-B(\Gamma_2,\Gamma_1)$ as given by
Eq.~(\ref{eq:B-Delta}), except for $\Gamma_1=\Gamma_2$, when
Eq.~(\ref{eq:B-Delta}) does not apply (we have only defined
$B(\Gamma_2,\Gamma_1)$ for $\Gamma_1\neq\Gamma_2$). We can calculate the
contribution from the cells
 $\Gamma_1=\Gamma_2$ to Eq.~(\ref{eq:Dm})
separately. We note that since $dN^a$ takes (most of the time) values 0 or 1,
$(dN^a)^2=dN^a$ and thus $\langle(dN^+-dN^-)^2\rangle=\langle dN^++dN^-\rangle$, or $\langle \delta
\Nnet^2\rangle = \langle\delta N_{\rm ch}\rangle$, up to
terms of order ${\cal O}(d\Gamma)^2$. Therefore, since $n=dN/d\Gamma$, the
integrand in Eq.~(\ref{eq:Dm}) for $\Gamma_1=\Gamma_2$ is $\langle
(n^{\rm net})^2\rangle/\langle n^{\rm ch}\rangle =
1/d\Gamma$. 
Summation over all cells with $\Gamma_1=\Gamma_2$ gives therefore a
contribution to $D_m$ equal to (up to infinitesimally small terms
${\cal O}(d\Gamma)$),
$4\,{\textstyle\left(\int d\Gamma\right)^{-1}}\! \int d\Gamma d\Gamma
\cdot 1/d\Gamma=4$. This is the value of $D_m$ for completely
uncorrelated particles. Adding the contributions from $\Gamma_1\neq
\Gamma_2$ we find therefore
\begin{multline}
  \label{eq:D-4-B}
  D_m = 4 \left( 1 - \,{\textstyle\left(\int d\Gamma\right)^{-1}}\!\int\! d\Gamma_1 \!\int\! d\Gamma_2\,
    B(\Gamma_2,\Gamma_1)\right) \\=4 \left( 1 - \int\! d(\Delta\Gamma) \,
    B(\Delta\Gamma)\right),
\end{multline}
where we used Eq.~(\ref{eq:B-Delta}) for the last equality.

\onecolumngrid

\end{document}